\documentclass[aps,prx,superscriptaddress,longbibliography,twocolumn]{revtex4-1}
\usepackage{amsmath}
\usepackage{amsfonts}
\usepackage{graphicx}
\usepackage{color}
\usepackage{dsfont}
\usepackage{verbatim}
\usepackage{bm}
\usepackage{mathtools}
\usepackage[normalem]{ulem}
\usepackage{comment}
\usepackage{multirow}

\begin{document}

\newcommand{\beq}{\begin{equation}}
\newcommand{\eeq}{\end{equation}}
\newcommand{\beqa}{\begin{eqnarray}}
\newcommand{\eeqa}{\end{eqnarray}}
\newcommand{\ben}{\begin{enumerate}}
\newcommand{\een}{\end{enumerate}}
\newcommand{\hs}{\hspace{1.5mm}}
\newcommand{\vs}{\vspace{0.5cm}}
\newcommand{\note}[1]{{\color{red}  #1}}
\newcommand{\notea}[1]{{\bf #1}}
\newcommand{\new}[1]{{#1}}
\newcommand{\ket}[1]{|#1 \rangle}
\newcommand{\bra}[1]{\langle #1|}
\newcommand{\im}{\dot{\imath}}
\newcommand{\tg}[1]{\textcolor{blue}{#1}}
\newcommand{\f}{f^{\phantom{\dagger}}}
\newcommand{\lmda}{{\bm{\lambda}}}
\newcommand{\Tr}{\text{Tr}}
\newcommand{\tX}{\tilde{X}}
\newcommand{\U}{\mathrm{U}}
\newcommand{\s}{\mathrm{S}}

\title{Quantum State Complexity in Computationally Tractable Quantum Circuits}

\author{Jason Iaconis}
\affiliation{Department of Physics and Center for Theory of Quantum Matter, University of Colorado, Boulder, Colorado 80309, USA}
\date{\today}

\begin{abstract}

Characterizing the quantum complexity of local random quantum circuits is a very
deep problem with implications to the seemingly disparate fields of
quantum information theory, quantum many-body physics and high energy physics.
While our theoretical understanding of these systems has progressed in recent
years, numerical approaches for studying these models remains severely limited.
In this paper, we discuss a special class of numerically tractable quantum
circuits, known as quantum automaton circuits, which may be particularly well
suited for this task.  
These are circuits which preserve the computational basis, yet can produce
highly entangled output wave functions.  Using ideas from quantum complexity
theory, especially those concerning unitary designs, we argue that automaton
wave functions have high {\it quantum state complexity}. We look at a wide
variety of metrics, including measurements of the output bit-string distribution
and characterization of the generalized entanglement properties of the quantum
state, and find that automaton wave functions closely approximate the behavior of
fully Haar random states. In addition to this, we identify the generalized
out-of-time ordered 2k-point correlation functions as a particularly useful
probe of complexity in automaton circuits. Using these correlators, we are able
to numerically study the growth of complexity well beyond the scrambling time
for very large systems. As a result, we are able to present evidence of a linear
growth of design complexity in local quantum circuits, consistent with
conjectures from quantum information theory.

\end{abstract}

\maketitle

\section{ Introduction}

Understanding the evolution of a quantum wave functions from a simple
initial state to a generic vector in an exponentially large  Hilbert space is a
notoriously difficult problem in modern theoretical physics. Aspects of this
evolution underlie important open problems in quantum information theory,
quantum many-body physics and high energy physics. Great progress has been made
in recent years by focusing on local random circuit models, which
provide a relatively clean system where these dynamics can be
studied~\cite{Nahum1, PhysRevX.8.021014,Nahum3,KhemaniVishHuse, Keyserlingk2}. 
A particularly important element of a generic quantum dynamics is the concept of
information scrambling.  
Originally studied in the context of black holes~\cite{HaydenPreskill,DonPage},
 scrambling defines the process whereby initially local information spreads
throughout the system and becomes
stored in the many-body non-local entanglement of the state. 
Similar works have since
used this concept to gain insight into how closed
quantum systems reach equilibrium and thermalize under a generic Hamiltonian
dynamics~\cite{mblarcmp}.

Two of the main tools which have been used to understand information scrambling
are the entanglement entropy of the quantum state and the evolution of the
out-of-time-ordered (OTO) correlation function.  It can be shown that the
entanglement entropy in these systems grows linearly with time until it
reaches a near maximal value~\cite{Nahum1}, and a decay of the out-of-time
ordered 4-point correlator has been shown to be equivalent to the
Hayden-Preskill definition of scrambling~\cite{RobertsYoshida2}. While such
measurements are useful, it has become clear that these relatively simple
measures cannot capture all the fine grained aspects of the random unitary
evolution. Two states may look maximally scrambled
according to these two measures and yet have important differences in the
precise way the information is stored non-locally.

Quantum state complexity theory has been suggested as a means to quantify these
differences~\cite{BrandaoPreskill,RobertsYoshida,LiuLloyd}. Roughly speaking,
the complexity of a quantum state is the depth of the \emph{smallest} local
unitary circuit which can create the state from an initial product state. 
In random circuit models, the growth of quantum state complexity directly
corresponds to an increased difficulty in distinguishing the pure quantum state
from the maximally  mixed state~\cite{BrandaoPreskill}. This is a physical
property whereby initially local information is more effectively hidden in high
complexity states.

It is known that a generic Haar random state will have a
complexity which is exponentially large in system size $N$. 
As a result, almost all quantum states cannot be efficiently simulated, even
with a quantum computer~\cite{Poulin_2011}.  A state which is the output of a
depth $D$ random circuit composed from a
universal gate set will have a complexity  which is conjectured to grow linearly
with $D$~\cite{Brown_2018,susskind2018black}. Ensembles of these wave functions
form what is known as an approximate projective unitary k-design~\cite{Gross_2007}. 
Measurements on k-designs can approximate, for large enough k, arbitrarily high moments of
measurements on fully Haar random states. 
On the other hand, states which are output from Clifford circuits in general
form only a unitary 2-design~\cite{zhu2016clifford}.  Although these wave
functions display volume law entanglement and information scrambling, they are
still of relatively low complexity and only approximate a few moments of
the Haar random states. 


In this paper, we show that high complexity quantum states can be prepared from a
special type of \emph{non-universal} local quantum circuit. These circuits,
which we call `automaton' quantum circuits, consist of any quantum gate which
preserves the computational basis. 
These automaton circuits have very recently started to be used as a tool for
studying dynamics in quantum systems~\cite{GopalakrishnanBahti,Sarang1,iaconis1}.
Specifically, in~\cite{iaconis1}, it was realized that the operator entanglement
and OTOC properties of such circuits appear to give results which are identical
to that of a generic chaotic dynamics.
We go beyond this and show that, when acting on initial product states not in
the computational basis, automaton circuits produce highly entangled wave
functions in which the quantum state complexity grows with circuit depth in the
same way as in universal local random circuits. 
Furthermore, the evolution of these wave functions can be efficiently simulated
classically using a quantum Monte Carlo algorithm which we describe.  This
may be appreciated in the context of several other results
in quantum information theory which demonstrate that the presence of
entanglement in a quantum state is not enough to show that a quantum algorithm
which simulates the state achieves a speedup over a classical
algorithm~\cite{Gottesman_Knill,Clifford1,VanDenNest}.  Our results imply that
complexity of the wave function is also not a sufficient condition for such
purposes. 

We do not attempt to provide a rigorous proof that automaton circuits output
states of high complexity. Instead, we characterize the complexity of the
automaton states using a series of measurements which were developed to probe the
fine-detailed structure of wave functions. We consider metrics such as the generalized
$k^{\text{th}}$ Renyi entropy~\cite{LiuLloyd,LiuLloydZhu} and the sampled output
bit-string distribution~\cite{BrandaoHarrow}, which can be used
to differentiate between high and low complexity states which both have near
maximal bipartite entanglement entropy. We also consider other measurements such
as the fluctuation of entanglement entropy and the level spacing distribution of
the entanglement spectrum. We will see that by these measures, the  
automaton wave functions behave like highly complex states. 

In a dynamical context, the generalized k-point OTO correlation functions
can describe the growth of quantum state complexity beyond
the scrambling time~\cite{RobertsYoshida}. Again, according to this metric, complexity in automaton
circuits appears to grow in the same way as in generic Haar random circuits. 
Further, using our efficient quantum Monte Carlo algorithm, we are able to
numerically study the growth of these OTO correlation functions in
this poorly understood ``beyond scrambling regime" for very large circuits. By
doing this, we are able to identify specific k-point OTO correlation functions
which appear to track the precise rate of complexity growth in local random circuits and
give results which are consistent with the linear growth conjectured in the
literature~\cite{Brown_2018,BrandaoPreskill}. 

The rest of this paper is organized as follows. In section II, we will introduce
and describe key properties of the quantum automaton circuits. We also describe
the quantum Monte Carlo algorithm we use to simulate these wave
functions.
In section III, we review the concept of quantum state complexity, and  describe several measurements 
which we use to distinguish between high and low complexity states. We will see
that by these metrics, automaton states behave like high complexity Haar random
states. We contrast these results to those of low complexity Clifford wave
functions.  In section IV, we discuss the generalized k-point
out-of-time-ordered correlator as a probe of complexity growth in dynamic
systems. We will see that automaton circuits can make use of these
correlation functions to give us new insights into complexity growth beyond scrambling in local quantum
circuits.  In section V we summarize our results and discuss potential
applications of this work.

\section{Automaton Quantum Circuits}

\subsection{Definitions and Review of Previous Results}
In this paper, we define automaton dynamics simply as any unitary evolution of a
quantum system which does not generate any entanglement when applied to product
states in an appropriate basis (which we will choose to be the computational
basis). As stated in~\cite{iaconis1}, an automaton unitary operator $U$ acting on an
appropriate set of product states in a $d$-dimensional Hilbert space
- labeled $\ket{m}$, with $m \in \{0, \dots, d-1\}$ - permutes these states up
to a phase factor, i.e.
\begin{eqnarray}
U \ket{m} = e^{i \theta_m} \ket{\pi(m)} \,
\end{eqnarray}
where $\pi \in S_d$ is an element of the permutation group on $d$ elements.

Similar unitary circuits with sparse output distributions have been studied in the
quantum information literature, where it was shown that efficient classical
simulation methods exist~\cite{vandennest1,vandennest2}. 
These circuits were first studied in a condensed matter context in
\emph{integrable} models in~\cite{GopalakrishnanBahti} and~\cite{Sarang1}.
In~\cite{iaconis1}, it was realized that a generic automaton evolution leads to
dynamics which appear to show `quantum chaotic' behavior. The out-of-time
ordered correlators propagate ballistically and saturate to the consistent
values for a Haar scrambled operator.  While automaton circuits do not generate
entanglement in the computational basis, a key property is that they do
generically generate a high degree of operator entanglement.  That is, the
evolution
\begin{eqnarray}
\mathcal{O} \rightarrow U^\dagger \mathcal{O} U
\end{eqnarray}
can be very complex  and shows many of the generic features of a Haar random
unitary evolution. 

One important example of such an automaton gate is a quantum version of the
CCNOT gate
\begin{eqnarray}
CCNOT(\theta)_{123} = 1 - \Pi_{12} - \Pi_{12}e^{i \theta}X_3.
\end{eqnarray}  
When $\theta=0$, this is the classical Toffoli gate which is known to be
universal for \emph{classical} reversible computation and can therefore implement any
permutation $\pi \in S_d$ on the computational basis states $\ket{m}$, $m \in
\{0,\dots d-1\}$. When $\theta \neq 0$,
such a gate also includes a state dependent phase.

A second important automaton gate set is the set
\begin{eqnarray}
\{\, \text{CNOT}, \text{SWAP}, R_z(\theta) \, \},
\end{eqnarray}
where $R_z(\theta) = e^{i \theta Z}$ implements a single qubit rotation about
the Z axis. At $\theta=\pi/2$, all three gates belong to the Clifford group.
The set of Clifford gates is capable of generating volume law entanglement when applied to
an appropriate initial product state, and the dynamics can be exactly simulated
classically \cite{Gottesman_Knill,Clifford1}. Therefore, the
automaton gate set generalizes the above Clifford group by allowing single qubit
rotations by arbitrary angles.

Note that both sets of gates defined above are universal for \emph{quantum}
computation if supplemented by any single qubit gate which does not preserve the
computational basis \cite{toffoli_universal}. 

We first review a few important analytic results derived in \cite{iaconis1}, in the case that
the automaton circuit is composed of $CCNOT(\theta=0)$. First, an
initially local diagonal operator $\mathcal{O}_{diag}$, will evolve into a
superposition over $O(d)$ other diagonal operators (where $d=2^N$ for qubits) and will have a
near maximal average operator entanglement.  Second, initially off-diagonal operators
will evolve into all elements of the conjugacy class of $S_d$, which implies
that an initial operator can evolve into $O(d^d)$ possible
off-diagonal operators. That is, a generic operator can evolve, under automaton
dynamics, into a \emph{super-exponential} number of other possible operators. Finally,
the recurrence time of a quantum evolution describes the time it takes for an
initial wave function to return to a nearby quantum state so that
$\bra{\psi_0}U\ket{\psi_0} \sim O(1)$.  For automaton circuits, the
recurrence time of an initial state (not necessarily a product state in the
computation basis) corresponds to the order
of a random element of the permutation group $S_d$ and on average gives $t_{rec}
\rightarrow exp(\lambda \sqrt{d/\log(d)})$ as $d \rightarrow \infty$.  

We also note that in \cite{iaconis1} it was found that the operator spreading in
automaton circuits,
as quantified by the 4-point out-of-time-ordered correlation function, behaves
identically to that of a Haar random chaotic circuit. In particular, the
operator weights spreads ballistically with a wave front which broadens
with a power law which is consistent with the universal exponents of a generic local chaotic
dynamics~\cite{PhysRevX.8.021014}.
 
In what follows, we take a complementary approach and study the evolution of
quantum \emph{states} which are initially product states in a basis orthogonal
to the computational basis. We will refer to the output
of such circuits as automaton wave functions. This approach allows us to focus
on the entanglement and complexity of the resulting wave function, and lets us
compare our algorithm with known variational Monte Carlo techniques.

\subsection{A Variational Monte Carlo Algorithm}

The defining feature of automaton circuits, that computational basis states
only evolve to other computational basis states, is what allows us to simulate
automaton wave functions on a classical computer. Despite their apparent
simplicity, such an evolution produces highly nontrivial wave functions when
applied to initial wave functions which are not product states in the
computational basis. 

We start with an initial ansatz wave function
\begin{eqnarray}
\ket{\psi_0} = \sum_m c_m \ket{m},
\end{eqnarray}
where we assume we know the coefficients $c_m$ exactly. Throughout
this paper, we will often choose $\ket{\psi_0}$ to be a product state in the
$X$ basis, $c_m = (-1)^{{\bf m}\cdot \bm{\sigma} }/d$, where ${\bf m}$ is a
binary vector representation of the integer $m$, and $\bf{\sigma}$ is a vector
of Pauli $X_i$ eigenvalues of $\ket{\psi_0}$. However, this need not be
the case, and we can choose any initial state $\ket{\psi_0}$ for which we have a
variational ansatz $c_m$.

We then time evolve the wave function by applying the quantum circuit
\begin{eqnarray}
U_{\lmda} = \prod_{t=1}^T \left [ \left (\prod_j U_{j,j+1}^{(t)} \right ) \left( \prod_j U_{j+1,j+2}^{(t)}\right) \right]
\end{eqnarray}
where $\lmda$ are the variational parameters which represent the precise set of
gates $\{ U_{j,j+1}^t\}$ which are applied. The resulting wave function is then
\begin{eqnarray}
\ket{\psi(t)} = U_\lmda \ket{\psi_0} = \sum_m c_m e^{i \theta_m}
\ket{\pi_\lmda(m)}.
\end{eqnarray}
Again $\pi_{\lmda}(m)$, is the permutation on the computational basis states,
$\ket{m}$, which is implemented by $U_\lmda$.
Therefore, we can exactly calculate the coefficients of the final wave function
$\ket{\psi(t)} = \sum_x \psi_\lmda(x,t) \ket{x} $  as
\begin{eqnarray}
\psi_\lmda(x,t) = \bra{x}\psi(t)\rangle = c_{\pi_\lmda^{-1}(x)} \exp[i
\theta_{\pi_{\bm{\lambda}}^{-1}(x)}].
\end{eqnarray} 
For a circuit with $N$ qubits and depth $T$, this can be calculated in a time
which scales like $O(NT)$. That is, since $\ket{m}$ only evolves to a
simple product state, $\ket{\pi(m)}$, instead of a
superposition over basis states, we can simply classically sample the initial
states $\ket{m}$ and track their time evolution. Nevertheless, as long as
$\ket{\psi_0}$ is not a product state in the computational basis,
$\ket{\psi(t)}$ will generally evolve into a volume law entangled state. In this way
we are able to classically simulate the circuit evolution of highly entangled
quantum wave functions in a way which is equivalent to the well known
variational Monte Carlo methods. 

We can therefore efficiently calculate estimates for simple operator expectation
values as
\begin{eqnarray}
\langle \mathcal{O} \rangle &=& \bra{\psi_0} U^\dagger \mathcal{O} U
\ket{\psi_0} \nonumber \\ 
	&=& \sum_{xy} c_y^* c_x^{\phantom{*}} \bigg \langle y \bigg | \bigg
(\prod_t U_t^\dagger \bigg) \mathcal{O}
	\bigg ( \prod_t U_t \bigg ) \bigg | x \bigg \rangle \label{eqn:circ} \\
&=& \sum_{x,y} \psi^*_{\bm{\lambda}}(\pi(y),t) \psi_{\bm{\lambda}}(\pi(x),t)\,
o\big(\pi(x),\pi(y)\big).
\end{eqnarray}

Since $U$ is an automaton circuit, then if $\mathcal{O}$ is a simple Pauli
operator we have $o(\pi(x),\pi(y)) = f(\pi(x)) \delta(y,x^\prime)$ with 
\begin{eqnarray}
x^\prime &=& \pi_{\bm{\lambda}}^{-1}(\pi_\mathcal{O}(\pi_{\bm{\lambda}}(x)).
\end{eqnarray}
Therefore, we can write 
\begin{eqnarray}
\langle \mathcal{O} \rangle &\approx& \frac{1}{M} \sum_{x_i=1}^M
\psi^*_{\bm{\lambda}}(\pi(x_i^\prime),t)
\psi_{\bm{\lambda}}(\pi(x_i),t) \, f(x_i) 
\\ &=& \frac{1}{M} \sum_{x_i=1}^M c_{x_i^\prime}^* c_{x_i}^{\phantom{*}} e^{i\left(\theta_{x_{i^{\phantom{\prime}}}}-\theta_{x_i^\prime}\right)} \,f(x_i)
  .
\end{eqnarray}

Note that for a generic off-diagonal operator $\mathcal{O}$, to determine which
state $\ket{x^\prime(t)}$ has a nonzero overlap with $\mathcal{O}\ket{x(t)}$, we
perform the full forward and backward time evolution in Eq.~(\ref{eqn:circ}) for
each time step independently. Estimating the full time dependence of
$\mathcal{O}(t)$ therefore takes a time $O(NT^2)$. On the other hand,
if $\mathcal{O}$ is a diagonal operator, then $x^\prime = x$ and we can get an
estimate for the entire time evolution in a time that scales like $O(NT)$. 

We finally note that using this approach to studying quantum circuit dynamics
allows us to make use of other tools developed in the context of variational Monte
Carlo algorithms. For example, one may incorporate Jastrow factors \cite{PhysRevB.71.241103},
Lanczos steps \cite{Becca_2015, PhysRevB.64.024512} or other perturbative corrections to the quantum wave
function \cite{Iaconis_2018}.  Furthermore, a promising direction for future work may
involve applying automaton circuits to RBM or other neural network wave
functions. Such models were studied for a subset of automaton gates in
\cite{jonsson2018}.

\begin{figure}[t]
\includegraphics[scale=0.6]{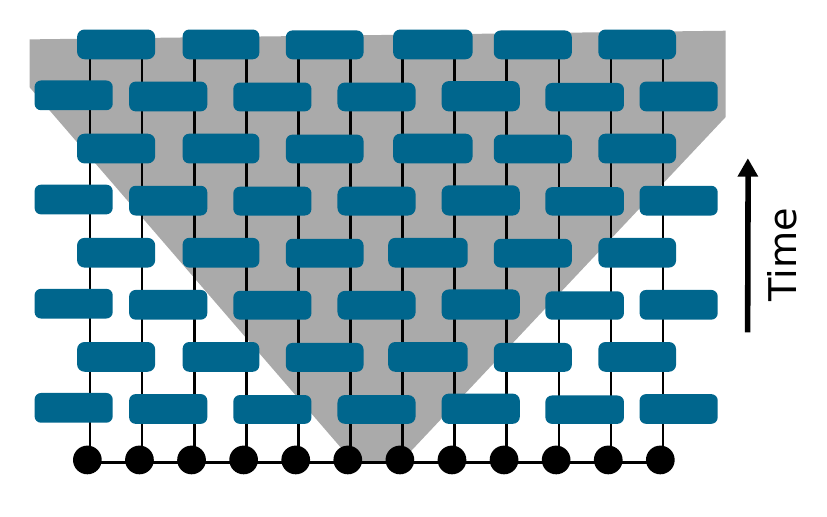} \\ 
\includegraphics[scale=0.7]{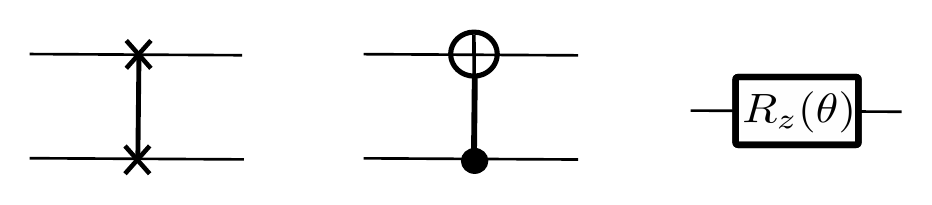}
\caption{The local random circuit architecture used throughout this paper. Each
2-site gate is chosen randomly to be one of 3 basic automaton gates: the SWAP
gate, the CNOT gate or the single site rotation about the z-axis
$R_z(\theta)=e^{i\theta\hat{Z}}$ (applied
independently to each site with a random angle $\theta$).}
\label{fig:circ}
\end{figure}

In the rest of this work, we focus on a specific 1D random circuit model
consisting of two site gates in alternating layers as shown in
Fig.~\ref{fig:circ}. The gates in this circuit are randomly chosen to be
either the two site SWAP or CNOT gates or a single site rotation by a random
angle $\theta$, $R_z(\theta)=e^{i\theta \hat{Z}}$.
We also compare the results to those of a random Clifford circuit, where we
randomly choose the gates to be either the two site SWAP or CNOT gates or the
single site Hadamard gate.

\section{Quantum State Complexity}

\subsection{Background}

Quantum complexity theory quantifies the difficulty of
particular tasks for a quantum computer, in terms of
the minimum number of basic quantum gates a computation requires.
Interestingly, in contrast to classical complexity theory, in the quantum
setting one can also meaningfully discuss the complexity of a quantum
\emph{state}. Roughly speaking, the complexity of a quantum state is the size of
the smallest k-local quantum circuit required to prepare the state from an initial
simple reference state. Unlike with classical bit-strings, creating a given
quantum state from a given initial state may in general require an exponentially
long quantum circuit. In fact, since the number of possible quantum circuits is
exponential in gate number, while the number of quantum states is
super-exponential in system size, one can show that {\it almost all} wave
functions require an exponentially long circuit to prepare.  
 
Importantly, the quantum state complexity of a wave function can be directly
related to measurable physical properties. 
This can be seen in the strong notion of complexity put
forward in reference \cite{BrandaoPreskill}.  In their work, the authors define
the complexity of a quantum state, $\ket{\psi}$, as the size of the smallest
local circuit, $U$, which when combined with measurement, $M$ in the
computational basis can distinguish $\ket{\psi}$ from the maximally mixed state
$\rho = \frac{1}{d}\mathbb{I}$. Mathematically, we define
\begin{eqnarray}
\beta_r = && \quad\, \max_{M} \, \quad \text{Tr}\left [ M(\ket{\psi} \bra{\psi}
\, - \, \rho_0) \right] \nonumber \\ && \text{subject to}\,\,\, M \in M_r(d)
\end{eqnarray}
where $M_r(d)$ is the set of generalized measurements composed of a unitary
circuit of depth $r$ acting on a Hilbert space of size $d$, which is followed
by a projective measurement in the computational basis.  We say that
$\ket{\psi}$ has strong
$\delta$-state complexity less than $r$,  $\mathcal{C}_{\delta}(\ket{\psi}) < r$ 
if
\begin{eqnarray}
\beta_r \ge 1-\frac{1}{d} - \delta.
\end{eqnarray} 
This is a very useful operational definition of complexity. It is directly
related to an experimental property of $\ket{\psi}$, the probability of
distinguishing $\ket{\psi}$ from the maximally mixed state with some fidelity
$(1-\delta)$, given a measurement implemented on a circuit of size at most $r$.
As $\delta \rightarrow 0$, this definition of complexity implies the weaker
condition, that $\ket{\psi}$ requires a minimum circuit of depth $r$ to be
prepared, but the converse is not in general true.

Theoretically, complexity in random unitary circuits can be understood using
another important concept, namely that of {\it unitary designs}
\cite{Gross_2007}.  An ensemble of quantum gates $\varepsilon = \{p_i, U_i\}$
acting on $\mathcal{H}^{\otimes d}$ is said to form an approximate unitary
k-design if the average over all such operators approximates the first $k$
moments of the Haar measure on all $d$ dimensional unitary operators.

A similar concept applies to ensembles of quantum states.  An ensemble $\nu$ of
pure states, $\psi$, form a complex projective k-design if
\begin{eqnarray}
\mathbb{E}_{\nu}[p(\psi)] = \int_{\nu_{_{Haar}}} d \psi \, p(\psi) \quad \forall \, p \, \in \, 
Hom_{(k,k)}(\mathbb{C}^d)
\end{eqnarray}
where $p$ is the space of polynomials homogeneous of degree $k$ both in the
coordinates of vectors in $\mathbb{C}^d$ and in their complex conjugates
\cite{LiuLloydZhu}.  In other words, for a complex projective k-design, all
expectation values which can be written as a polynomial of degree $k$ in the
wave function coefficients must be equal to the expectation value of a random
quantum state chosen from the Haar measure.

These two seemingly different ideas, complexity and design, are in fact very
closely related.  Since almost all states in the Hilbert space have
exponentially high complexity, one may guess that relatively high complexity states are
required to approximate distributions on the Haar measure. 
In reference \cite{BrandaoPreskill}, such a rigorous connection is made between
unitary designs and quantum state complexity. It was shown that a unitary
k-design has, with high probability, a complexity $\approx O(N k)$.
More precisely, it was shown that for a k-design in a $d=q^N$ dimensional
Hilbert space formed from a set of $|G|$ basic gates, that 
\begin{eqnarray}
\hspace{-3mm} Pr[C_\delta(\ket{\psi}) \le r] \le 2(1+\epsilon) d N^r |G|^r \left (\frac{16
k^2}{d(1-\delta)^2}\right)^k,
\end{eqnarray}
which qualitatively remains very small until $r \approx k(N-2\log(k))/log(N)$. 
In other words, with high probability, such a k-design has state complexity at
least $O( k N/log(N))$. 

Unitary k-designs define a fine-grained hierarchy of
quantum states of increasing complexity. This concept is referred to in the
literature as {\it complexity by design} and is explored, for example, in
\cite{BrandaoPreskill, RobertsYoshida} and \cite{LiuLloyd}.

This idea allows us to bridge the gap between 
local universal unitary gates, which form the basis of local quantum circuits, and generic
$d$-dimensional unitary operators $U$ which a random circuit tries to emulate.
Characterizing the rate of complexity growth in local random circuits is an
important open question. 
In \cite{BrandaoHarrow}, it was shown that, with high probability, a local random circuit
composed of universal gates of depth $O(N k^{11})$ forms at least a unitary
$k$-design. In other words, the design order of a local random circuit grows
polynomially with circuit depth. It is expected, however, that this bound is not
very tight. In \cite{Brown_2018}, it is argued that the average complexity of local
circuits in fact grows linearly with circuit depth. 

Conversely, there are certain ensembles of quantum gates which are known to form only a fixed finite k-design. 
The set of Pauli strings, $\s = \otimes_{i=1}^N \sigma_i^{\alpha_i}$, forms an exact 1-design.  The set of Clifford gates on
$q$-dimensional qudits are known to form a unitary 2-design in general, a
3-design for $q=2$ and never form a 4-design \cite{zhu2016clifford}. While wave functions resulting
from  Clifford circuits are sufficient to see properties such as volume law entanglement
and information scrambling, we will see that there exists a range of observable
properties which they do not possess and which are characteristic of the higher
complexity regime. 
In a sense, quantum state complexity generalizes the notion of
information scrambling. The degree to which information is spread non-locally
in a quantum state can be quantified by the difficultly of recovering such
information.

In the rest of this section, we proceed in the following way. We first identify several
observable properties of quantum states which have been explored in the
literature and can be used to diagnose complexity beyond scrambling. 
Strict bounds on these measurements can be formulated when they are averaged over
a unitary design. 
We will measure these properties in automaton wave functions. The results
suggest that automaton wave functions have high state complexity.  Where useful,
we will also compare these measurements to those of Clifford circuits, which are known
to form a finite low-order unitary design. 
As a consequence, we will show that while a universal local
gate set is sufficient for creating wave functions of high complexity, it is not
in fact necessary. Indeed, wave functions of high complexity can be formed by
acting with an automaton circuit, and therefore such a circuit evolution can be
simulated efficiently with a classical computer in the manner described in the
previous section.  We note, however, that the specific measurements used in the
rest of this section cannot generally be implemented efficiently with a Monte
Carlo algorithm and so we instead simulate the exact automaton and Clifford
dynamics on relatively small system sizes. In section \ref{sec:otoc}, we will
study measures of complexity which can be efficiently implemented using our
Monte Carlo algorithm.

\subsection{Deviations from the Maximally Mixed State}

Like normal random variables, fluctuations in the matrix elements of random
unitaries must satisfy strict bounds. For fully Haar random unitaries, these
bounds imply that probability amplitudes of randomly sampled bit-strings follow
the well known `Porter-Thomas' distribution, $p(x_j) = |\langle
x_j|\psi\rangle|^2 \sim d e^{-d p(x_j)}$. Such an output distribution is a signature of
quantum chaos, and sampling random bit-strings from this distribution for
universal local random gates is expected to be a hard problem to simulate classically
\cite{Boixo_2018}. 
 
If a unitary matrix $U$ is drawn instead only from a unitary
k-design, fluctuations of matrix elements can be shown \cite{BrandaoHarrow} to satisfy a weaker
bound. In this case, one finds that for any two unit vectors $\ket{\alpha}$ and
$\ket{\beta}$
\begin{eqnarray}
\Pr_U \left [\bra{\beta} U \ket{\alpha} \, \ge \, \frac{\gamma}{d} \right ] \le
(1+\epsilon)e^{-\min(k,\gamma)}.
\label{eqn:bound}
\end{eqnarray} 
If we let $\ket{\alpha} = \ket{\psi_0}$ and $\ket{\beta}$ be any basis vector,
this bounds the fluctuations of the coefficients $|c_n|^2$ of
$\ket{\psi(t)}$. Indeed, if we let $k \gg N$, as we expect for a universal
local random circuit at late times, and assume the fluctuations saturate this bound, we see
that k-designs approximate the Porter-Thomas distribution arbitrary well for
sufficiently large $k$.

For automaton gates, the bit-string distribution in the computational basis
remains constant. Therefore, for an initial state orthogonal to the
computational basis, sampling the computational basis bit-strings is equivalent to sampling
from the maximally mixed state. However, we find that bit-strings measured in the
orthogonal `X' basis form a nontrivial probability distribution and further this
distribution satisfies the strict bounds set for generic unitary k-designs. 

\begin{figure}
\includegraphics[scale=0.5]{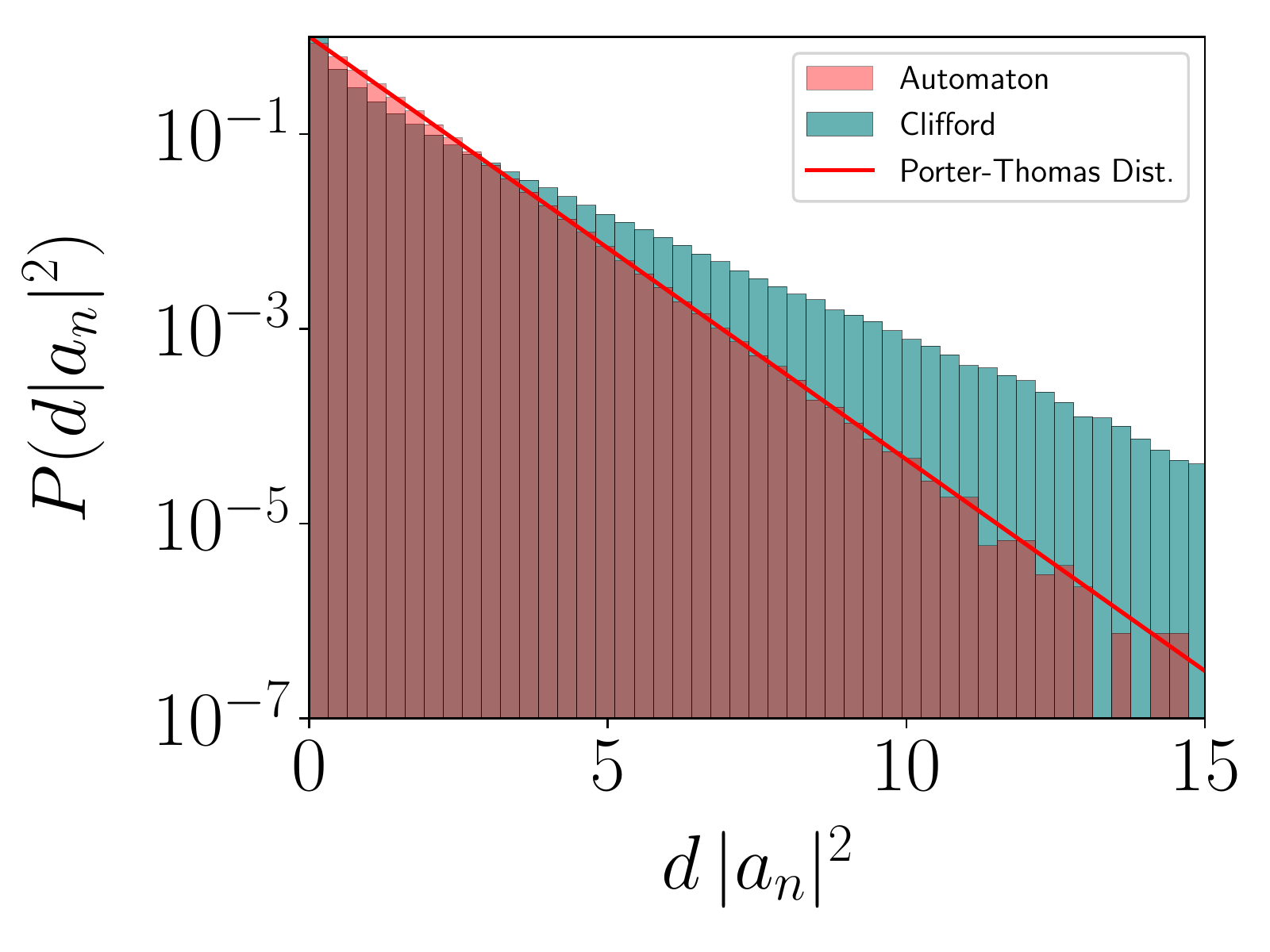}
\caption{The probability distribution of bit-strings $P(2^N |a_n|^2)$ as
measured in the $x$-basis, for automaton and
Clifford wave functions on $N=16$ sites. The fluctuations  of bit-string amplitudes
in the automaton wave functions obey the strict bound for unitary $\gamma$
designs given by Eq.~(\ref{eqn:bound}) up to at least $\gamma \sim N$. The Clifford wave function, on the
other hand, only obeys this bound up to $\gamma\sim 3$.} \label{fig:bitstring}
\end{figure}

To see this, we simulated the exact quantum circuit dynamics of an initial
product state with all spins oriented perpendicular to the computational basis,
with gates chosen randomly from our automaton gate set. To compare, we also
simulated random Clifford circuits, with gates chosen randomly from
$\{CNOT,SWAP,H\}$ and acting on a random initial product state.  
We should emphasize that since we are interested in the
distribution of a many-bit output, we cannot use the polynomial time classical
algorithm to simulate either the automaton or Clifford
circuits~\cite{Jozsa2013ClassicalSC}. Instead, we are forced to track the
evolution of the entire wave function for a small system
size.  We simulated a circuit with $L=16$ sites and circuit depth $D=100$. 
A histogram of the final projective measurement outcomes for both cases is
shown in Fig.~\ref{fig:bitstring}.  For the automaton circuit, the state is
initialized with all spins oriented perpendicular to the computational basis,
and the final output bit-strings are measured in the x-basis.  The results are
averaged over 100 different circuit realizations.

We see that the probability of different basis strings decays exponentially, up
to the resolution we are able to measure. For  
the Clifford circuits, the Porter-Thomas bound is satisfied only up to $\gamma=3$, but
is violated for $\gamma>3$. The implication is that for automaton
circuits, measurements in the orthogonal basis are extremely uniform in the same
way as for high complexity Haar random states.

\subsection{Entanglement and Complexity}

The pattern of entanglement in quantum states is very closely related to the
quantum state complexity.  We will see that the entanglement in states drawn
from a unitary k-design must satisfy certain constraints. As shown by Page,
nearly all quantum states chosen from the Haar measure will have a nearly
maximal amount of entanglement \cite{DonPage}. More
precisely, the bipartite von Neumann entanglement entropy of a random quantum state with
Hilbert space $\mathcal{H} = \mathcal{H}_A\otimes\mathcal{H}_B$ can be shown
to be  
\begin{eqnarray}
S_{vN} \ge \log(d_A) - \frac{1}{2\ln(2)} \frac{d_A}{d_B} 
\end{eqnarray}  
where $d_A \le d_B$ are the dimensions of $\mathcal{H}_A$ and $\mathcal{H}_B$
respectively.

Our definition of quantum state complexity implies that high
complexity states cannot easily be  distinguished from the maximally mixed
state. This property necessarily requires the state to be nearly maximally
entangled, so that the reduced density matrix $\rho_A$ is close to the maximally
mixed state for all subregions $|A|<\frac{L}{2}$. Therefore, the process of scrambling requires that
initially local information becomes stored in the non-local many body
entanglement of the wave function. However, the converse statement is not always
true. States of high entanglement are not necessarily always of high complexity.
To distinguish between states of different complexity, we need to develop more
fine grained measures of entanglement.

\begin{figure}[t]
\hspace{-70mm} {\bf a)}  \\ 
\includegraphics[scale=0.40]{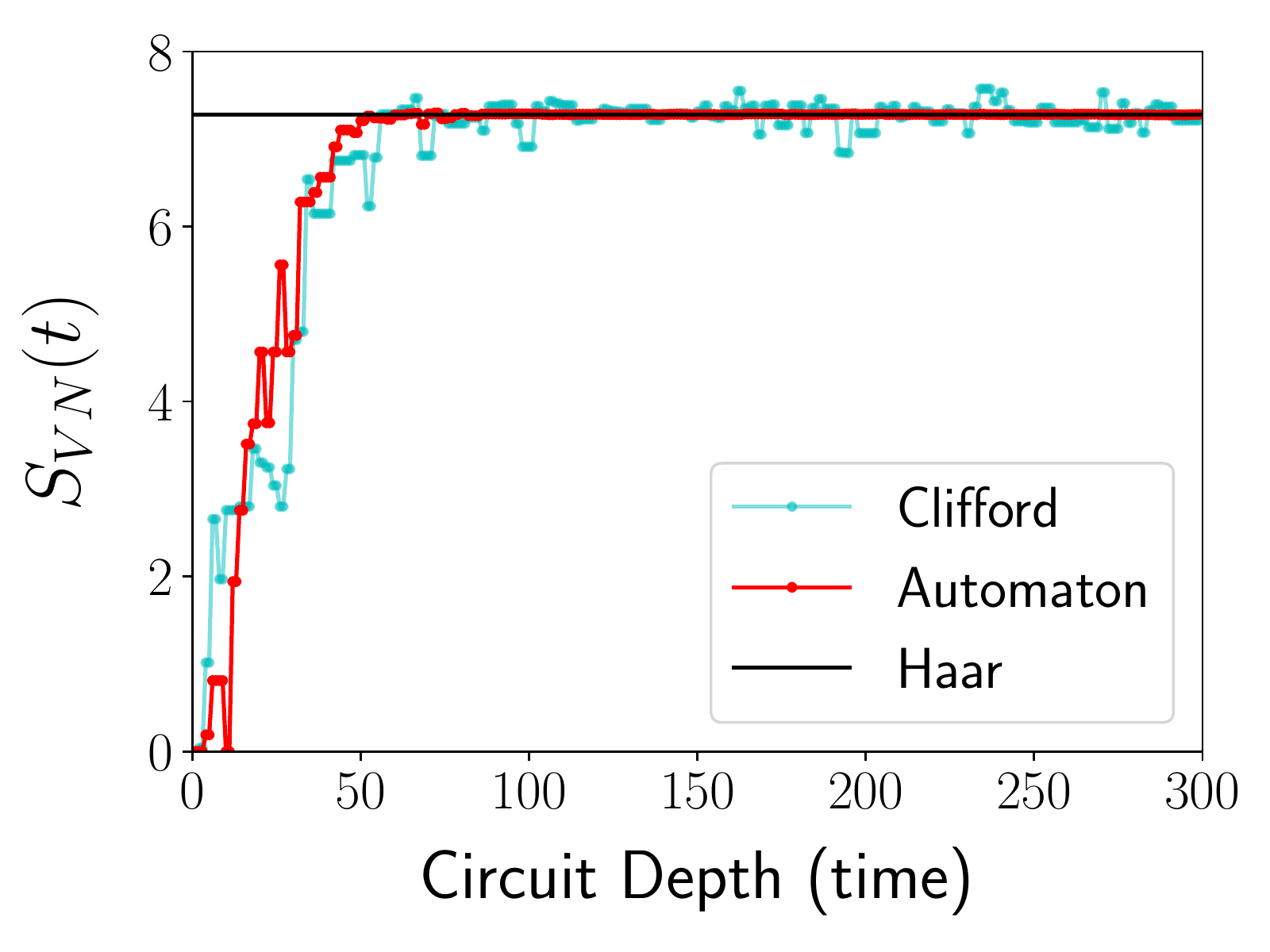} \\
\vspace{1pc}
\hspace{-70mm} {\bf b)} \\
\vspace{1pc}
 \hspace{-10mm} \includegraphics[width=0.38\textwidth,height=0.37\textwidth]{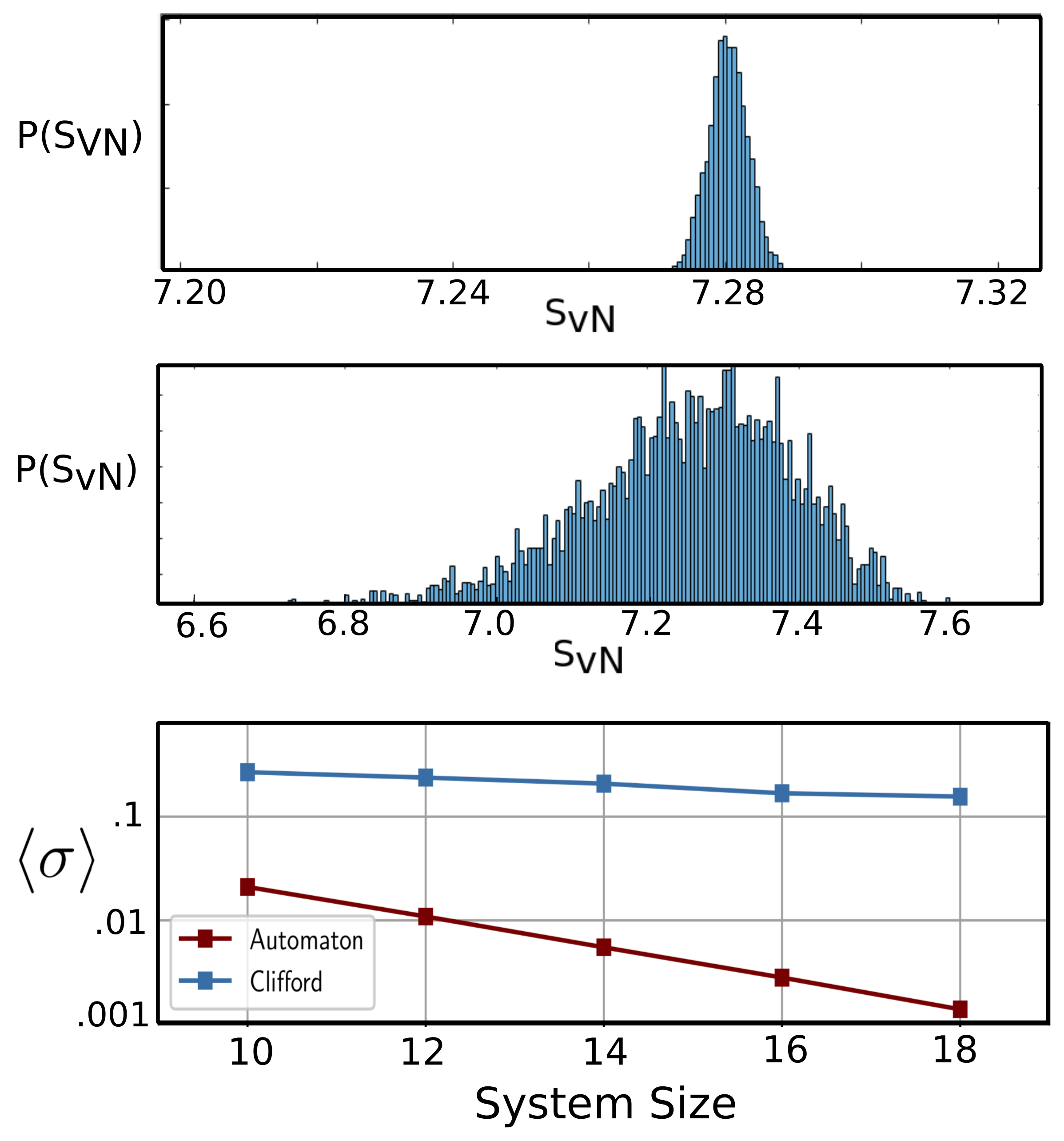}
\caption{{\bf a)} The bipartite entanglement entropy $S_{vN}$ as a function of time
for both Clifford and automaton circuits. In both cases, the late time
entanglement averages to the Haar random `Page entropy', however the temporal
fluctuations are significant in the Clifford circuit, while they appear
negligibly small in the automaton circuit.  {\bf b)} This uniformity of entanglement
can be seen in a single realization of an automaton wave function, if we
measure the entanglement in all possible bipartitions of the lattice. We show
the probability distribution of the entanglement entropy across the different
partitions for the automaton {\it (top)} and Clifford {\it (middle)} wave
functions. The standard deviation of these distributions {\it (bottom)},
$\langle\sigma\rangle = \sqrt{\langle (S_{vN}-\langle S_{vN}\rangle)^2\rangle}$, decays
exponentially with system size for automaton circuits and appears to saturate to
a constant value for Clifford circuits. 
}\label{fig:EE} 
\end{figure}

In Fig.~\ref{fig:EE} a), we show the time evolution of the bipartite von Neumann
entanglement entropy $S_{vN}(t)$ for a single circuit realization of both
automaton and Clifford circuit types.  In both cases, we observe a short regime
of linear entanglement growth followed by a late time regime where the
entanglement saturates near the volume law Page value $S_{vN} = L/2 -
1/(2\ln(2))$. The main difference between the two cases is that for the
automaton circuit, after reaching saturation, $S_{vN}$ remains very close the
exact Page value at all times, while in the Clifford circuit there are
relatively large fluctuations in $S_{vN}(t)$. We will argue that these
fluctuations in the entanglement entropy are a sign that a state is not drawn
from a sufficiently high unitary design. 

$S_{vN}$ measures the entropy of the reduced
density matrix, $\rho_A$, which encodes all information about observables which
can be measured locally in region A.  Fluctuations in $S_{vN}(t)$ therefore imply there
are fluctuations in the value of some measurement in region A. 
In Brandao et. al. \cite{BrandaoPreskill}, it was shown that for a unitary k-design,
the higher order moments of a generic expectation value are bounded by
\begin{eqnarray}
\mathbb{E}_{\ket{\psi}}\left [ \left ( \text{Tr}(M \ket{\psi}\bra{\psi}) -
\mathbb{E}_{\ket{\psi}} [ \text{Tr}(M \ket{\psi}\bra{\psi})] \right )^k \right ]
\le \left ( \frac{k^2}{d}\right )^{k/2}.\nonumber \\
 \label{eqn:fluct}
\end{eqnarray} 
For a highly complex state, which forms a large k unitary design, the higher order
fluctuations on all measurements become very small.
If we partition our lattice into regions $A$ and $B$, and let
$M$ be any projective measurement implemented on the spins in subsystem $A$,
then this should also bound fluctuations of the entanglement entropy. 
Therefore, the temporal \emph{fluctuations} in the entanglement entropy are
evidence that the Clifford circuit is of lower complexity than the automaton
circuit.  

Using this intuition, we can develop an entanglement measure which acts on a
quantum wave function at a single time and quantifies the degree of the entanglement fluctuations.
This measure is simply the full probability distribution of bipartite entanglement entropies
measured across all $N \choose N/2$ bipartitions of the lattice. 
Comparing the entropy across many different lattice partitions effectively 
measures the multi-partite entanglement of the wave function
\cite{RevModPhys.80.517}, similar to the entanglement measure developed by Meyer
and Wallach \cite{Meyer_2002}.  

 We show the histogram of these entropies in Fig.~\ref{fig:EE}b) for both
automaton and Clifford circuits. We see that indeed, for automaton circuits,
almost all bipartitions of the state have the same entanglement entropy, which is
very close to the Page entropy. However, for Clifford circuits, while the
average entanglement entropy is equal to the Page entropy, there are
significant, $O(1)$,
variations in this measurement depending on which bipartition is selected.
This implies that the Clifford states are much less uniform than the automaton wave
functions. Therefore, the variance in measurements in automaton states should
satisfy Eq.~\ref{eqn:fluct} for a much higher value of $k$ compared to Clifford
states.

Perhaps the most direct connection between entanglement and unitary design can be made by
studying the generalized Renyi entanglement entropies.  In \cite{LiuLloyd}, it
was shown that the higher order $\alpha$-th Renyi entropies can be used as a
direct probe of the design order.  The $\alpha$-Renyi entropy is defined as
\begin{eqnarray}
\hspace{-3mm} S^\alpha(\rho_A) = \frac{1}{1-\alpha} \log(\Tr[\rho_A^\alpha]) = \frac{1}{1-\alpha}
\log \left [\sum_i \lambda_i^\alpha \right], 
\end{eqnarray}
where $\lambda_i$ are the eigenvalues of the reduced density matrix $\rho_A$. As
$\alpha \rightarrow \infty$, $S^\alpha(\rho_A)= S_{\min}(\rho_A) =
-\log(\lambda_{\max})$, approaches the min entropy. $S_{\min}(\rho_A)$ simply probes
the largest eigenvalue of $\rho_A$, and bounds all other Renyi entropies
$S^\alpha(\rho_A) \le S_{\min}(\rho_A), \forall \alpha$. In \cite{LiuLloyd}, it was
shown that the $\alpha$-Renyi entropy averaged over a unitary $\alpha$-design is
nearly maximal. Therefore, the higher-order Renyi entropies can be seen as a
probe of higher order complexity in the wave function. It was shown that
\begin{eqnarray}
\mathbb{E}_{\nu_k} \left [ S^k(\rho_A) \right ] \ge d_A + \mathcal{O}(1).
\end{eqnarray}
where $\mathbb{E}_{\nu_k}$ is the average over the k-design distribution of unitary matrices. 
Furthermore, they showed that
\begin{eqnarray}
\mathbb{E}_{\nu_k}\left [ \Tr\{\rho_A^k\} \right ] = \mathbb{E}_{\text{Haar}}
\left[ \Tr \{\rho_A^k \} \right ].
\end{eqnarray} 
In other words, the $k-th$ Renyi entropies are all nearly maximal up to an
$\mathcal{O}(1)$ constant for a unitary k-design, and the trace of $\rho_A^k$ 
exactly equals the Haar random value. This exact equality does not hold in
general for the Renyi entropies since the log of an average does not in general
equal the average of a log. 

We measure the different Renyi entropies for Haar random local circuits,
automaton circuits and Clifford circuits. In all cases, we again perform the
measurements on small circuits where we can track the evolution exactly. In
principle, we could measure these Renyi entropies for larger automaton circuits
using our classical algorithm. However, this involves measuring higher order
`Swap' operators, which have a value which is exponentially small in the amount
of entanglement. Since the amount of entanglement of a bipartition grows like a
volume law, this measurement becomes exponentially hard in these systems.

\begin{figure}
\includegraphics[scale=0.45]{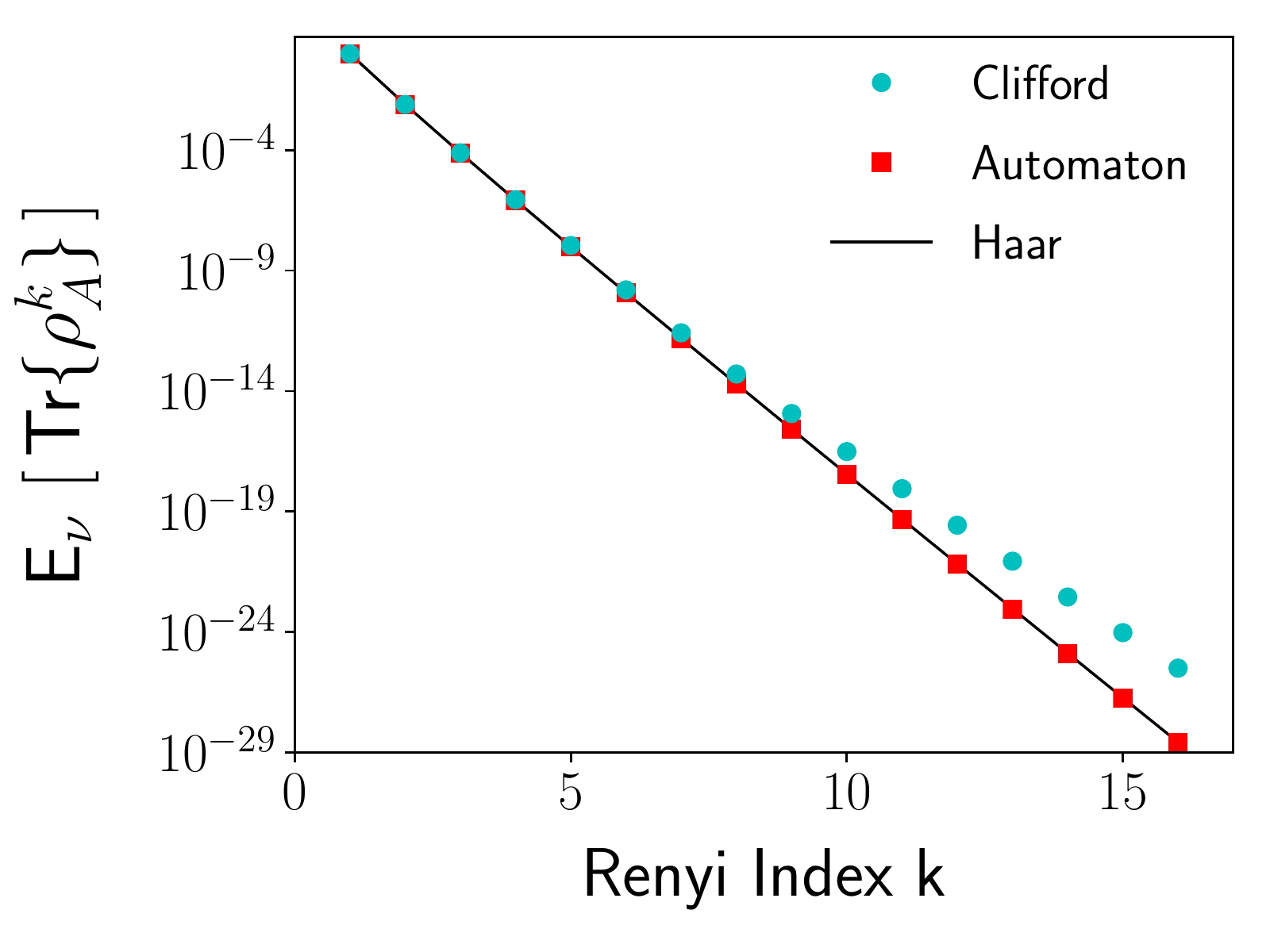}
\caption{ The average trace of $\rho_A^k$ for different values of the Renyi
index k. We find that the expectation over circuits of $\text{Tr}(\rho_A^k)$, is
equal to the Haar random value for automaton circuits for all $k$ that we
tested. For Clifford circuits, which form a unitary 3-design,
$\mathbb{E}[\text{Tr}(\rho_A^k)]$ are only constrained to match the Haar random value up to
$k=3$, and show significant deviation above $k\approx 5$.}\label{fig:renyi} 
\end{figure}

In Fig.~\ref{fig:renyi}, we show the expectation value for the
k$^{\text{th}}$ Renyi entropy, as measured in both automaton and Clifford
circuits. Amazingly, the expectation value for the automaton wave functions is
exactly equal to the Haar random value for all values of $k$ which we measured.
For Clifford circuits, on the other hand, the expectation value matches the Haar
value for low Renyi index, but deviates significantly at higher value of $k$.
$\mathbb{E}_\nu[\rho_A^k]$ is the expectation value of the k$^{\text{th}}$
Renyi entropy measured over the ensemble of circuits $\nu$. At high Renyi index,
$k$, small fluctuations away from this mean will be amplified. Therefore, these
results are again consistent with the hypothesis that fluctuations of random
measurements are highly suppressed in automaton wave functions, to the extent
that such measurements mimic that of a fully Haar random wave function.
Interestingly, in
\cite{LiuLloydZhu}, it was found that the infinite order Renyi entropy
$S_{\infty} \sim -\log(|\lambda_{max}|)$, saturates near its maximal value after only
an $O(N)$ time. Such a state is known as `max scrambled'. Although
the complexity of the quantum state continues to grow past the max scrambling
time, all max scrambled states will appear maximally complex according to the
Renyi entanglement measures. Our results strongly imply that automaton wave
functions will become max scrambled for polynomial depth circuits.

\subsection{Entanglement Spectrum}

Entanglement spectrum is the name given to the statistical distribution of the
eigenvalues of a reduced density matrix \cite{PhysRevLett.101.010504,es}. 
The \emph{spacing} between these eigenvalues form a distribution which is known for
different ensembles of random matrices \cite{mehta2004random} and generically follows a Wigner-Dyson
distribution. For a random $U(N)$ unitary matrix, the spacing between
eigenvalues follows the Gaussian unitary ensemble (GUE).  These Wigner-Dyson
distributions have the special property that there is repulsion between
neighboring eigenvalues. On the other hand, the reduced density matrix of  wave
functions that result from integrable dynamics do not, in general, form a
random matrix. In such a case, the eigenvalues of $\rho_A$ do not show the same
degree of level repulsion and may follow a simple Poisson distribution.

To measure the entanglement spectrum, we first rewrite the wave function $\ket{\psi}$ using
the Schmidt decomposition.
\begin{eqnarray}
\ket{\psi} = \sum_i \lambda_i \ket{\alpha_i} \ket{\beta_i}
\end{eqnarray}
where $\lambda_i$ are real positive numbers. 

We can then define the entanglement spacing, $s_i =
\lambda_{i+1}^2-\lambda_{i}^2$, where we order the Schmidt coefficients such that
$\lambda_0 \le \lambda_1 \le \cdots \le \lambda_M$. For convenience
\cite{PhysRevLett.110.084101,PhysRevB.75.155111}, we define the level spacing
ratio 
\begin{eqnarray}
r_i = \min \left [ \frac{s_i}{s_{i+1}}, \frac{s_{i+1}}{s_i} \right].
\end{eqnarray}
The entanglement spectrum is then the probability distribution of the $r_i$
random variables.

\begin{figure}
\includegraphics[scale=0.50]{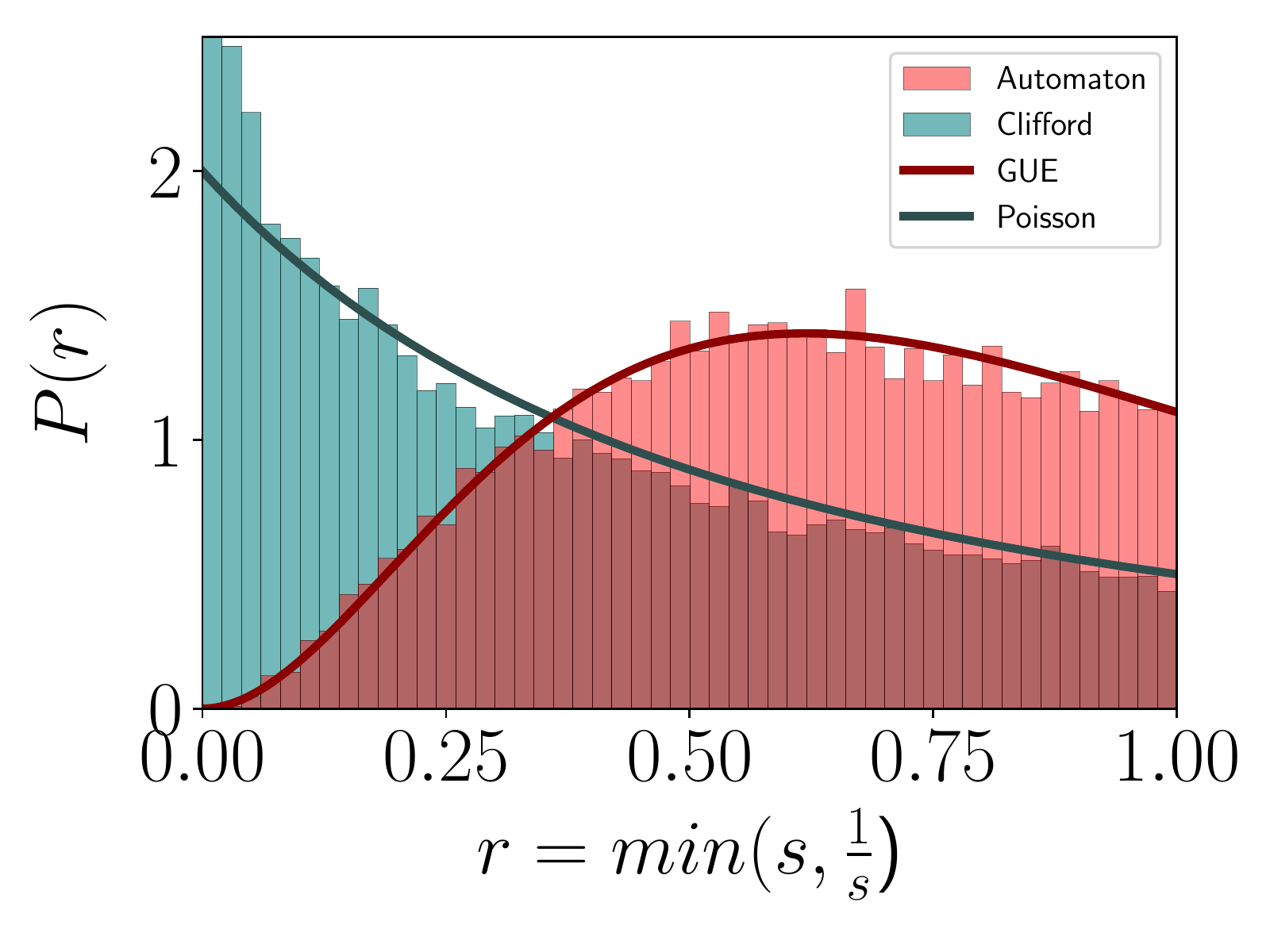} 
\caption{The level spacing distribution of the entanglement spectrum for
automaton wave functions show Wigner-Dyson GUE statistics, while the Clifford states show
Poisson like statistics. Wigner-Dyson statistics are expected for the 
eigenvalue distribution of random matrices and are a signature of quantum chaos.}
\label{fig:ES}
\end{figure}

In Fig.~\ref{fig:ES}, we show the entanglement spectrum statistics for wave
functions that result from both the automaton circuit and Clifford circuits. We
see that the spectrum in the automaton case follows very closely the universal
form of the Gaussian unitary ensemble
\cite{PhysRevLett.110.084101,PhysRevLett.122.180601}, while the Clifford states
do not show the same level repulsion and appear to follow a Poisson
distribution.

The relationship between chaotic dynamics and entanglement spectrum has been
studied in \cite{Shaffer_2014, Chamon_2014, Yang_2017}. However, a complete
theoretical understanding of the connection between quantum state complexity and
the entanglement spectrum is still lacking.  In \cite{Shaffer_2014}, it was
noted that dynamics under a universal set of quantum gates is sufficient to
generate $GUE$ statistics of the entanglement spectrum, while evolution under
Clifford gates results in Poisson statistics.  Here, we have shown that this
condition of a universal set of quantum gates is not necessary to generate GUE
statistics.  Indeed, we have created a state with such statistics using only the
automaton gate set, which can be simulated classically in the way outlined in
section II. It is interesting that such signatures of quantum chaos also appear
in wave functions which can be simulated classically.

It remains an open question whether there is a concrete relationship between
entanglement statistics and unitary k-designs.

\section{Measuring Complexity in Automaton Circuits} \label{sec:otoc}

\subsection{Generalized Out-Of-Time-Ordered Correlators} 

Out-of-time-ordered correlators (OTOCs) have recently been found to be an
important tool for characterizing operator spreading in quantum circuits. 
The 4-point OTOC
\begin{eqnarray}
\langle OTOC^{(4)} \rangle = \langle A(t)B(0)A(t)B(0) \rangle \label{4point}
\end{eqnarray}
measures the average degree of non-locality of an operator $A(t) =
U^\dagger A U$, and has been extensively studied in the context of
thermalization and quantum chaos \cite{Chaos_Qi,PhysRevX.8.021014,Maldacena_2016}. 
This quantity will only be small if $A(t)$ evolves into a highly non-local
operator.  In \cite{RobertsYoshida2}, it was shown that the
information-theoretic definition of scrambling is implied by the generic decay
of this four point function. Furthermore, any initial product state which is
evolved by such a unitary can be shown to be nearly maximally entangled.
 
Following the work of Roberts and Yoshida \cite{RobertsYoshida}, we can generalize this
operator and define the 2k-point out-of-time ordered correlators:
 \begin{eqnarray}
\langle OTOC^{(2k)} \rangle = \langle A_1(t)B_1(0) \dots A_k(t)B_k(0)
\rangle. 
\end{eqnarray}
Deep connections have been found between the generic
smallness of the 2k point functions, unitary k-designs and quantum circuit
complexity. A generic 2k-point function contains k copies of $U$ and k copies of
$U^\dagger$. Therefore, if $U$ is sampled from a unitary k-design then the
average of the 2k-point function over the ensemble $\{U\}$ must equal the
Haar random value, and therefore will be exponentially small. 
Since the four-point OTOC expectation value is quadratic in the $U$ and $U^\dagger$
operators, we see that only a unitary 2-design is necessary for scrambling.
We know, however, that the complexity of the wave function will
continue to grow well past this scrambling time.

These higher order correlators are therefore an important tool for understanding
complexity beyond scrambling in quantum dynamics. Crucially, the
generalized OTOC functions give us a probe which is insensitive
to the onset of lower-order forms of complexity. For example, the 4-point
function in local circuits takes on an $O(1)$ value throughout the `thermalization'
regime, before decaying to an exponentially small value after a time 
$t^*\sim O(L)$ for local circuits. This is in contrast to the
entanglement entropies, which can also be used to diagnose scrambling and complexity, but
which always require an exponentially hard measurement to implement. 
This feature of the OTOCs is both useful experimentally, and critically, allows
us to use automaton circuits to numerically probe the onset of complexity
efficiently in large systems.

We can therefore use these higher order correlation functions to probe the
structure of the wave functions output from quantum circuits. If we can find a 2k-point OTOC which
is nonzero, this implies that the unitary ensemble is not a k-design and the
wave function is likely of lower complexity. 

In \cite{RobertsYoshida}, it was shown that the average value of the
2k-point correlation function can directly give a lower bound for the quantum
circuit complexity of a unitary ensemble $\{ U \} = \varepsilon$,
\begin{eqnarray}
 C(\varepsilon) &\ge& (2k-1)2N \nonumber \\  
&& - \log\left [ \sum_{A_1\dots B_1\dots} \langle A_1(t)B_1
\dots A_k(t)B_k \rangle \right ].
\end{eqnarray}
This expression is useful for showing that there is a direct connection
between the generic decay of the higher-order OTOCs and circuit complexity.
Unfortunately, it is not very useful for numerically calculating a bound on
circuit complexity since the main contribution comes from calculating a sum over
an exponentially large number of operators each of which is in general
exponentially small. As we will explain below, in this work, we take an
alternative route by identifying special structured OTO correlators which have
an $O(1)$ value for low complexity dynamics. 

We will proceed as follows. We first identify a class of k-order OTOCs which
have a special recursive structure which can be physically motivated and can be
easily generalized. We then also perform a brute force search over all 2k-point OTOCs for a
fixed value of k. These correlators are not as easily generalizable, but give
a more complete picture of complexity growth in local random circuits.  In
both cases, we are able to efficiently measure the correlation function in high
depth automaton circuits with a large number sites.  We find that in both cases 
the correlators eventually decay to an exponentially small value in automaton
circuits, providing strong evidence in very large systems that automaton
circuits produce high complexity wave functions. Further, our brute force
search is able to identify a large, linear in $k$, regime where the quantum wave function
appears scrambled but the higher order OTOCs have not yet decayed. This
gives us an unprecedented ability to numerically study complexity
growth in local quantum circuits.

\subsection{Recursive K-point Functions}

We begin by studying a special instructive class of k-point OTOC functions
which often retain an $O(1)$ expectation value beyond the scrambling
time $t_{sc}$. 
In these correlators, the time evolved Heisenberg operators
$\tilde{\mathcal{O}} = \U^\dagger \mathcal{O} \U$ can be treated as a
generalized unitary operator $\U^{(n)}$:
\begin{eqnarray}
\U^{(0)} &=& \U \\
\U_{\mathcal{O}}^{(1)} &=& \U^\dagger \mathcal{O} \U \\
\U_{\mathcal{O},\mathcal{O}^\prime}^{(2)} &=& \U^{(1)^\dagger}_{\mathcal{O}} \mathcal{O}^\prime
\U_{\mathcal{O}}^{(1)} \\
&& \dots \nonumber
\end{eqnarray}
The higher order OTOCs can be interpreted as assessing the scrambling properties of
$\U^{(n)}$. For example, for we can write 

\begin{eqnarray}
\left \langle \, \tilde{A} B \tilde{A} C \tilde{A} B \tilde{A} C^{\phantom{\dagger}} \right \rangle &=& 
\left \langle \U^{(1)\dagger }_{A} B \U^{(1)}_A \, C \, 
\U^{(1)\dagger}_{A} B \U^{(1)}_A \, C \right \rangle \nonumber \\
&=& \left \langle \,B(t) C B(t) C ^{\phantom{\dagger}} \right \rangle. 
\end{eqnarray}
where, following the notation of \cite{RobertsYoshida}, we let $\tilde{A} =
\U^\dagger A \U$. Therefore, this 8-point function under $\U$ can be thought of as a 4-point
function under $\U^{(1)}$.  These recursive OTOCs can always be interpreted as
4-point OTOCs with additional local operators hiding in the generalized
unitaries.

Under a fully Haar random $U(2^N)$ dynamics, all k-point
correlation functions will decay to an exponentially small value. Therefore,
not only does $\U$ have high quantum complexity, but so do the operators,
$\U^{(1)}=\U^\dagger A \U$, $\U^{(2)}=\U^{(1)\dagger}B \U^{(1)}$, etc. 

Conversely, for the known examples of exact unitary designs, such as the ensemble of Pauli
strings and Clifford circuits, the higher order generalized unitaries are of
lower complexity than the original operator. 

Consider the case where $\{\U\}$ is an ensemble of Clifford circuits. 
These are known to form a unitary 2-design in general, and a
3-design when the local Hilbert space is qubits, but never form a
4-design~\cite{zhu2016clifford}.  Therefore, when averaged over the ensemble of
all Clifford circuits, all 4-point functions are found to be exponentially
small, $\langle \tilde{A}B\tilde{A}B\rangle = 4^{-N}$.  However, the defining
feature of Clifford circuits is that they evolve Pauli strings to other Pauli
strings. Therefore, the generalized unitary operator $\U^{(1)}$ is simply a Pauli
string, $\U^{(1)} = \s = \otimes_i \sigma_i^{\alpha_i}$.  The ensemble of Pauli
strings $\{\s\}$ are known to merely form a 1-design, and so the 4-point
functions under $\{ \U^{(1)}\}$ do not decay to zero. The 4-point function under
$\U^{(1)}$ is an 8-point function under $\U$. Therefore, there always exist
8-point functions for Clifford circuits which do not decay to the Haar random
value.  Therefore, the fact that Clifford circuits {\it merely scramble} is
demonstrated by the fact that that while $\{\U\}$ scrambles, the ensemble of
unitary operators $\{A(t)=\U^\dagger A \U\}$ do not.
In this way, the higher order OTOCs expose a hierarchical structure of unitary
designs. 

With this understanding, we use the higher order OTOC to probe the
dynamics of automaton circuits in the `beyond scrambling' regime. 
Since automaton circuits apply nontrivial dynamics in the direction perpendicular
to the computational basis, we further define a set of recursive unitaries
which are composed only of single site $X$ Pauli operators:
\begin{eqnarray}
\U^{(0)} &=& \U \nonumber \\
\U_{i_1}^{(1)} &=& \U^\dagger X_{i_1} \U \nonumber \\
\U_{i_1i_2}^{(2)} &=& \U^{(1)\dagger}_{i_1} X_{i_2} \U^{(1)}_{i_1} \nonumber \\
\U_{i_1i_2\dots i_m}^{(m)} &=& \U^{(m-1)\dagger}_{i_1i_2\dots i_{m-1}}X_m
\U^{(m-1)}_{i_1i_2\dots i_{m-1}} \nonumber.
\end{eqnarray}

We then write down the special class of generalized OTOCS
\begin{eqnarray}
F^{(2^k)}_{i_1,\dots,i_{k-1},0} &=& \langle \U^{(k-1)\dagger}_{i_1\dots i_{k-1}}
X_0 \U^{(k-1)}_{i_1\dots
i_{k-1}}X_0 \rangle. \label{kpoint_rec}
\end{eqnarray}
Then, for example: 
\begin{eqnarray}
F^{(4)}_{L,0} &=& \langle \tX_L X_0 \tX_L X_0\rangle, \nonumber \\
F^{(8)}_{L,1,0} &=& \langle \tX_L X_1 \tX_L X_0 \tX_L X_1 \tX_L X_0 \rangle.
\label{kpoint_48}
\end{eqnarray}
$F^{(4)}_{L,0}$ is simply the usual 4-point OTOC which measures
operator scrambling, so that $F^{(4)}_{L,0}=1$ if and only if $[\tX_L,X_0]=0$.
On the other hand, $F^{(8)}_{L,1,0}$ measures the scrambling of $X_1$ under a time evolution by $\tX_L =
\U^\dagger X_L \U$. 
In this case, we have that $F^{(8)}_{L,1,0} = 1$ if either $[\tX_L,X_0]=0$ or $[\tX_L,X_1]=0$,
and therefore we have that 
\begin{eqnarray}
\mathbb{E}[F^{(8)}_{L,1,0}] \ge \mathbb{E}[F^{(4)}_{L,0}].
\end{eqnarray} 
These higher order OTOCs are a more strict measure of complexity, can
easily be generalized, and retain a simple interpretation as measuring the
scrambling properties of the generalized unitary operators.

\begin{figure}
\includegraphics[scale=0.5]{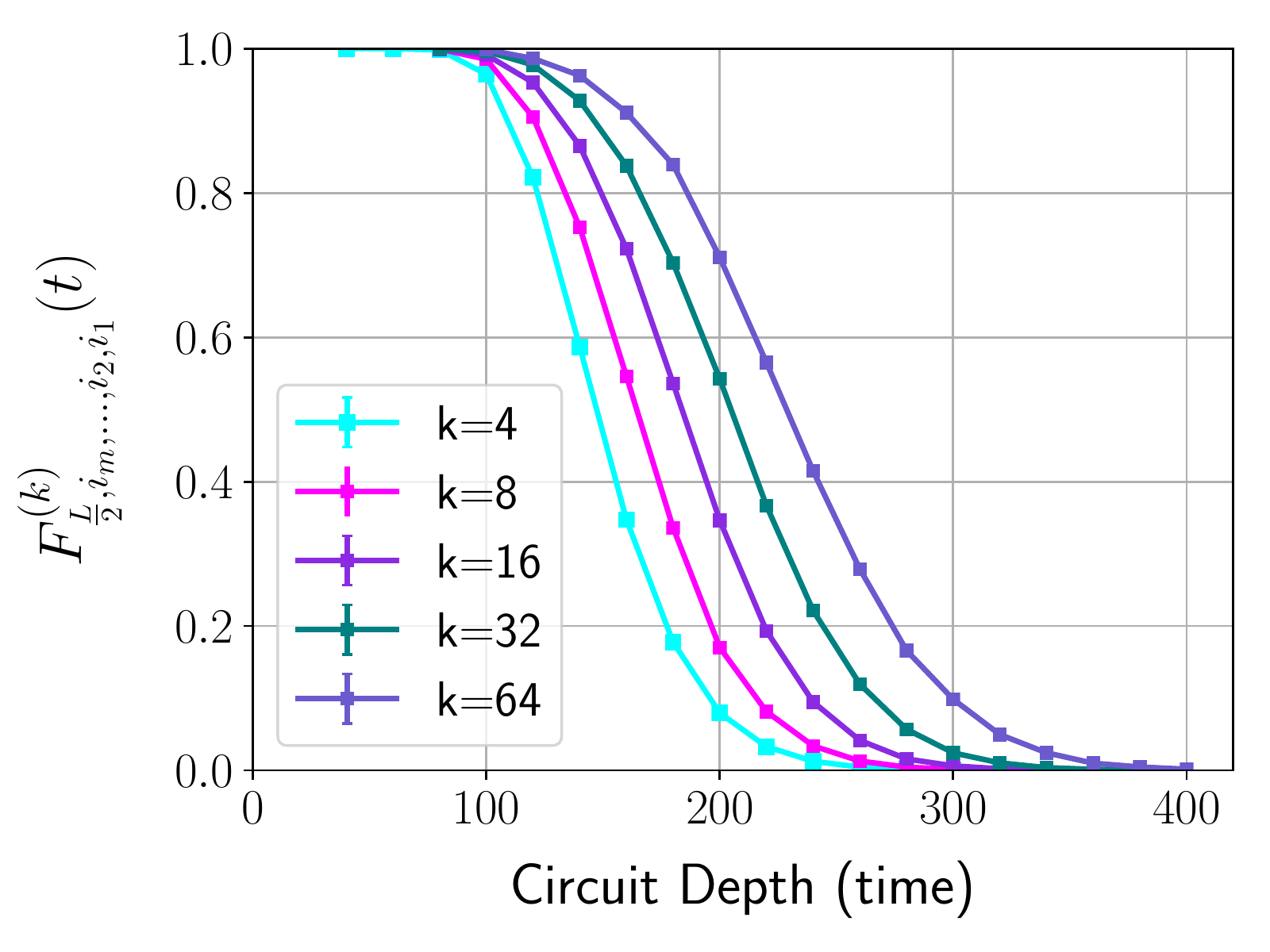}
\includegraphics[scale=0.5]{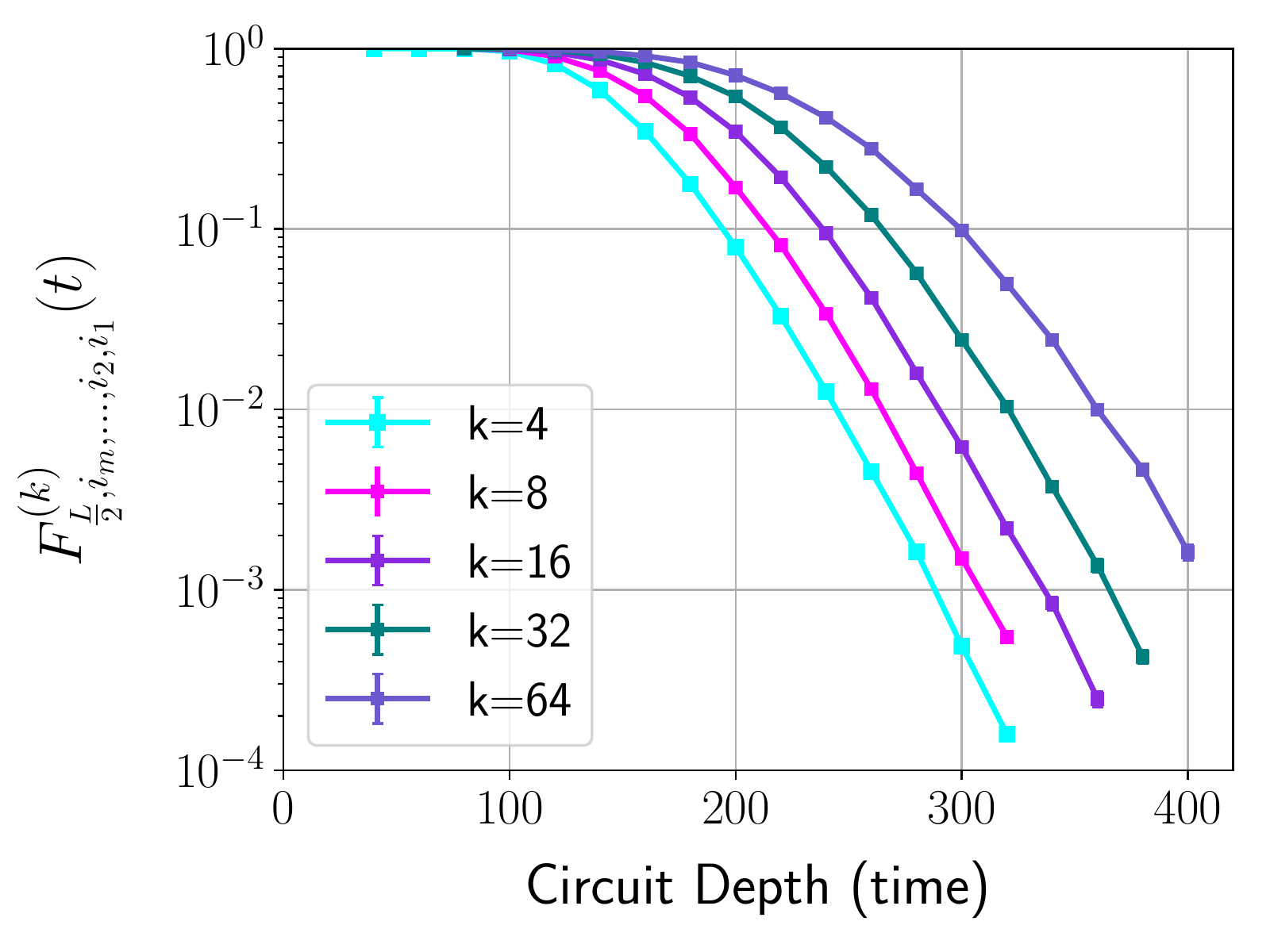}
\caption{The recursive k$^{\text{th}}$ order OTO correlators,
$\langle \text{F}_{\frac{L}{2},i_m,\dots,i_2,i_1}^{(k)} \rangle$, defined in
Eq.~(\ref{kpoint_rec}), for a system with $L=100$ sites with periodic boundary
conditions. These OTOCs are composed of single site Pauli $X$
operators acting on sites $\{\frac{L}{2},i_m, \dots,i_3,i_2,i_1\}=\{\frac{L}{2},
\log_2(k/2),\dots,2,1,0\}$.  See Eq.~(\ref{kpoint_48}), for the form of the
$k=4$ and $k=8$ operators.  We can measure these correlators for very high
values of k. {\it (Top)} The higher order OTOCs decay at progressively later
times. The highest order OTOCs, remain nonzero well past the scrambling time.
{\it (Bottom)} At late times, all correlators eventually decay at an
exponential rate.}\label{fig:otoc1}
\end{figure}

We measure these recursively defined operators and show the results in
Fig.~\ref{fig:otoc1}, for an automaton circuit with $L=100$ sites. The results
are again averaged over different random circuit realizations.
We point out several important features of this data.

First, we see that for automaton circuits, all generalized OTOC functions do
eventually decay to an exponentially small value. We take this as important
further evidence that automaton circuits have high quantum circuit complexity and the
resulting wave functions have a high quantum state complexity. 
Again, this should be  seen as a stark contrast to other examples of numerically
tractable quantum circuits such as Clifford circuits, for which we can always
find higher order OTOCs which do not decay at all.

Second, we see that in these circuits, the higher order OTOCs are nonzero at
later times than the usual 4-point function $F^{(4)}_{L,0}= \langle \tX_L X_0
\tX_L X_0 \rangle.$  This concretely demonstrates that in such local random
circuits there exists a well defined regime beyond the scrambling time where information
about the original state $\ket{\psi_0}$ is not completely lost. In these
`intermediate complexity states', local information from $\ket{\psi_0}$ can still be
probed using these special  measurements.  Further, note that the expectation
value of the higher order OTOCs at late times is generally much greater than
twice the
previous order, yet only requires twice the computational effort to measure. 

Finally, we see that the `scrambling time', $t^*$ for this class of higher order
OTOCs appears to only increase \emph{logarithmically} with order $k$. In
particular, we find that
\begin{eqnarray}
t^* = v_B L + v_k \log_2(k)
\end{eqnarray} 
In the next subsection, we will see that this is not a generic feature of the
higher order OTOCs.

\subsection{Higher Order OTOCs and Complexity}

A more complete picture of complexity growth in our local random automaton
circuits can be found by studying more general classes of k-point OTOCs. 

For local random circuits, it was shown in \cite{BrandaoHarrow}, that the
unitary design order will continue to grow far beyond the scrambling time.  It
is expected that
the complexity in these local circuits grows linearly with circuit depth, $D$,
for an exponentially long time \cite{Brown_2018}.  This linear complexity growth is of
great interest in the high-energy literature. In the context of the AdS/CFT
correspondence, the linearly growing complexity of the dual CFT is related to
the growth of an AdS wormhole \cite{Susskind_2016,PhysRevD.90.126007}.
In \cite{BrandaoHarrow}, a weaker result, that design order, (and
therefore the complexity), in these circuits grows at least like the polynomial
$D^{\frac{1}{11}}$, was proven rigorously.  Linear complexity growth can also be
rigorously shown in the limit of a large local Hilbert space dimension
\cite{BrandaoPreskill}. However, there is no known proof of linear complexity
growth for the more difficult case of a local random circuit composed of qubits.

\begin{figure}
\includegraphics[scale=0.52]{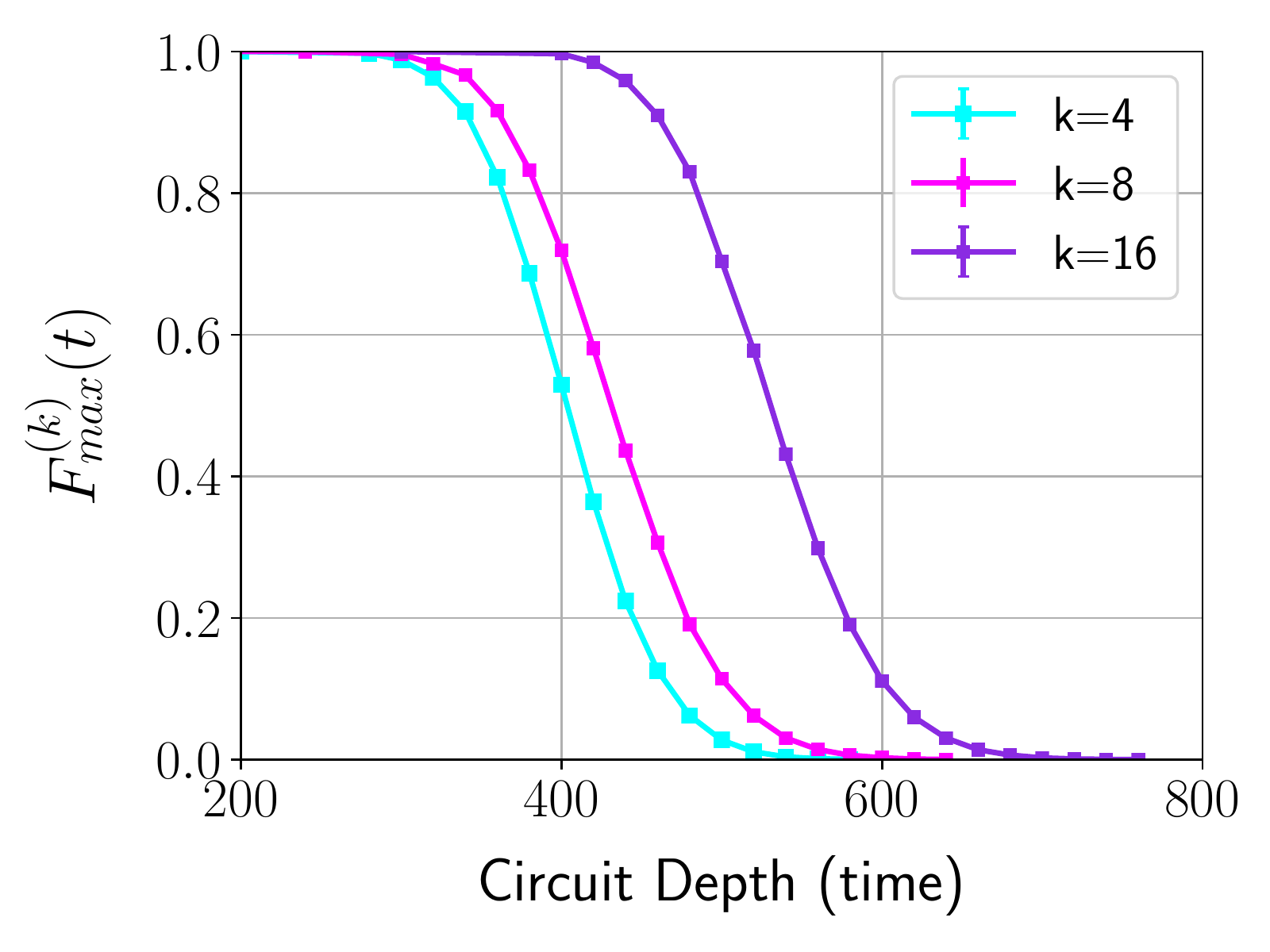}
\includegraphics[scale=0.52]{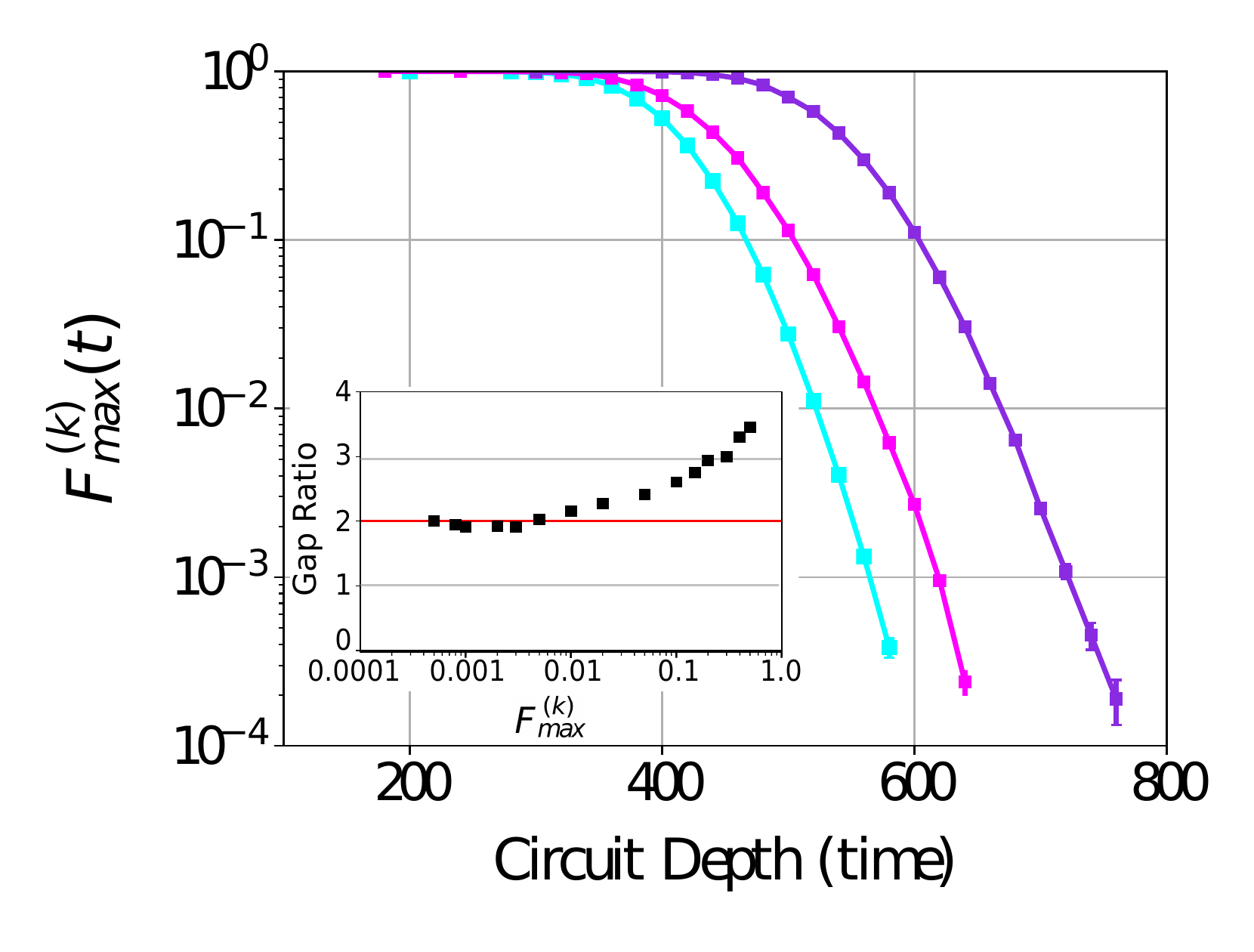}
\caption{A brute force search over a wide class of k$^{\text{th}}$ order OTOCs
gives the max OTOC $F_{max}^{(k)}$ (see text for a precise definition).
We see that the $k=16$ point function decays at a much later time in this case
compared to the recursive OTOCs of the previous section. In the bottom plot, we
show the same data on a logarithmic scale and define the scrambling time
$t^*_k(\epsilon)$ as the time beyond which $F_{max}^{(k)}<\epsilon$. In the
inset we see that the ratio between scrambling times $\Delta = (t^*_{16}-t^*_8)/(t^*_8-t^*_4)$ 
approaches $\Delta=2$ as $\epsilon \rightarrow 0$. }
\label{fig:otoc2}
\end{figure}

Quantum automaton circuits allow us to test this conjecture numerically in large
systems by using the generalized OTOC operators as a measure of complexity. To do this, we
should search over the full set of out-of-time-ordered operators
\begin{eqnarray}
OTOC_{max}^{(k)} = \max_{A_1,B_1,\dots,A_k,B_k} \mathbb{E}_\nu \left [\langle
\tilde{A}_1 B_1 \dots \tilde{A}_k B_k \rangle \right ]
\end{eqnarray}
where $A_i$, $B_i$ are any Pauli string operators. However, this
would require computing $4^{Nk}$ different expectation values, which is
clearly intractable even for small values of $N$ and $k$. 

Instead, we again restrict the operators $A_i$, $B_i$, to be only single site Pauli $X$
operators and define 
\begin{eqnarray}
F_{max}^{(k)} = \max_{i_1,\dots,i_k} \mathbb{E}_\nu \left [\langle \tX_{i_1} X_{i_2} \dots \tX_{i_{2k-1}}
X_{i_{2k}} \rangle \right ].
\end{eqnarray} 
In practice, we find a lower bound for this operator by searching only over
operators with support on a small fixed number of sites.
Even with these practical restrictions, $F_{max}^{(k)}(t)$ gives an upper bound on the design order of the
circuit at time $t$.  

For $k$ up to $k=16$, we search over all possible correlators
with $A_i,B_i \in \{X_0, X_1, X_{L/2-1},X_{L/2}\}$. 
For $k=4$, the maximum OTOC is the usual 4-point function $F_{max}^{(4)} =
\langle \tX_{L/2} X_0 \tX_{L/2} X_0 \rangle$. 
At $k=8$, we find the maximum expectation value occurs when the OTOC takes the
form
\begin{eqnarray}
F_{max}^{(8)} &=& \langle \tilde{\mathcal{O}}\tilde{\mathcal{O}}\rangle \\
\tilde{\mathcal{O}} &=& \tX_0 X_{\frac{L}{2}} \tX_0 X_0 
\end{eqnarray} 
To calculate $F_{max}^{(16)}$, we must measure $\sim32000$ different correlation
functions. In this case, we find that the maximum expectation value occurs for
\begin{eqnarray}
F_{max}^{(16)}&=& \langle \tilde{\mathcal{O}} \tilde{\mathcal{O}} \rangle \\
\tilde{\mathcal{O}} &=& \tX_0 X_{\frac{L}{2}} \tX_0 X_0 \tX_0 X_{\frac{L}{2}}
\tX_{\frac{L}{2}} X_0,
\end{eqnarray}
plus special equivalent permutations of these operators which are related by symmetry.

\begin{figure}
\includegraphics[scale=0.52]{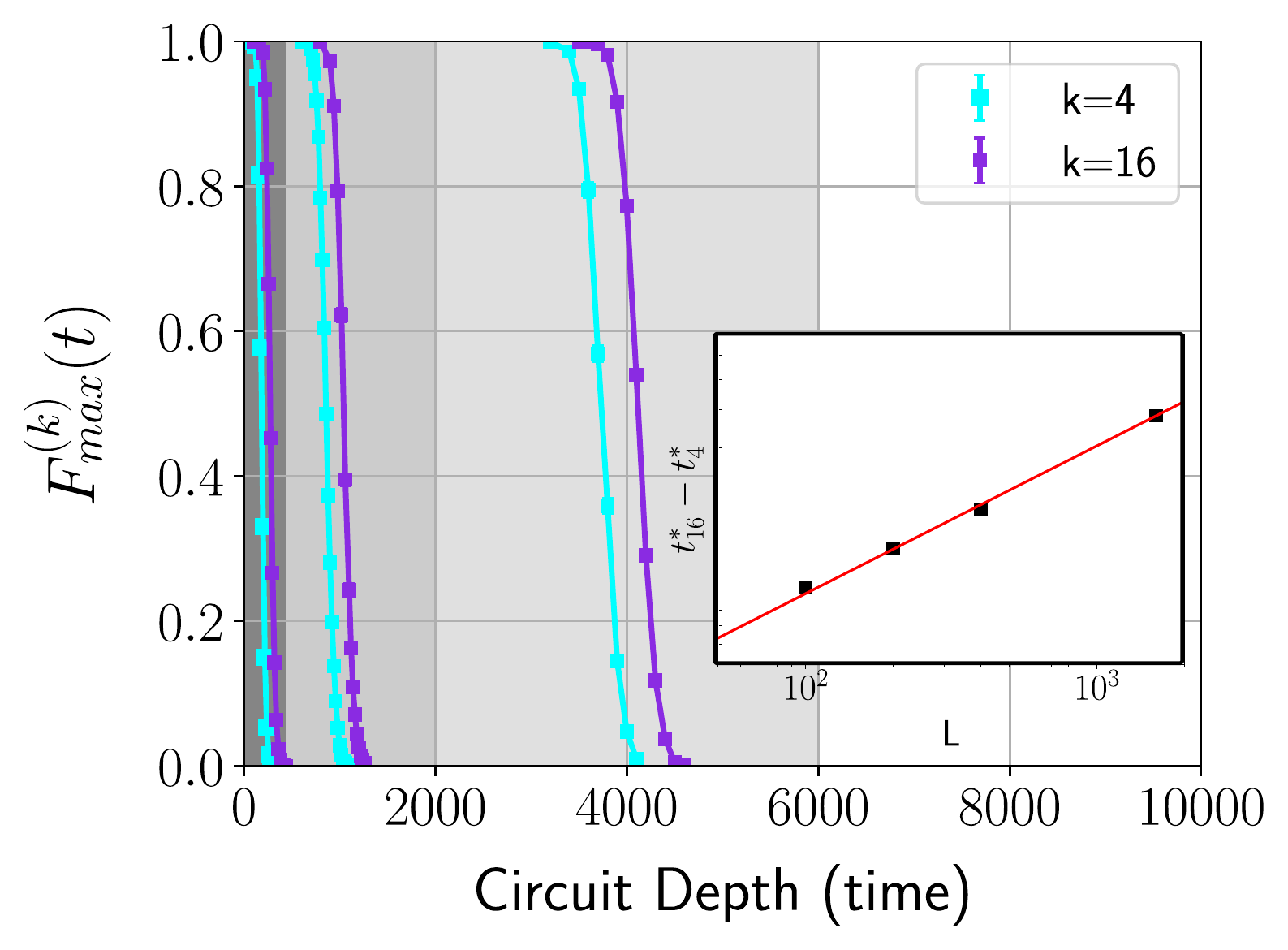}
\caption{$F_{max}^{(k)}(t)$ for $k=4$ and $k=16$ for different system sizes
$L=100, 400, 1600$ (Shown left to right in separate shaded regions). 
The gap between the $k=4$ and $k=16$ wave fronts grows with system size.
This gap size vs $L$ is shown in the inset to grow like $\Delta \sim L^\alpha$ with
$\alpha=0.48(1)$.
}\label{fig:otoc3}
\end{figure}

Note that we find this correlation function appears to `peak' only at special
values of $k$. That is, we find that at late times, $F_{max}^{(16)}(t) \gg
F_{max}^{(8)}(t)
\gg  F_{max}^{(4)}(t)$, but that this is not true for other values of $k$ which we tested
(up to at least k=24).
 
The results for the 4, 8 and 16 point functions are shown in
Fig.~\ref{fig:otoc2}. Again, we see that all correlation functions we
measured decay to an exponentially small value at late times.  The dramatic
difference is that the $k=16$ point function remains nonzero for a significantly
longer period of time. 

We see that the complexity growth in local random circuits is subject to the so
called switchback effect \cite{PhysRevD.90.126007}, whereby there is a delay in the onset of
linear complexity growth for initially local operators. This occurs due to the
exact cancellation of unitary gates outside the lightcone in the operator
evolution $\tilde{\mathcal{O}} = \U^\dagger \mathcal{O} \U$.
 
We define the time $t^*_k(\epsilon)$ as the circuit depth beyond which
$F^{(k)}_{max}<\epsilon$.  Due to the switchback effect, we expect that the linear
complexity growth begins only after we reach the scrambling time $t_{sc} = t^*_4$, when
the wave function forms an approximate 2-design.  At some later time $t^*_8$, we
will have $F_{max}^{(8)}<\epsilon$, and the wave function will form an
approximate 4-design. This difference $\Delta = t^*_8-t^*_4$ defines the rate of
complexity growth.  The inset in Fig.~\ref{fig:otoc2} b)  shows that
$t^*_{16}-t^*_8 \approx 2 \Delta$ for sufficiently small $\epsilon$.  
In Fig.~\ref{fig:otoc3}, we find that as the system size in increased, the size of the
`complexity gap' $\Delta$ grows like $\sqrt(L)$.  

Therefore, at least for these 3 values of $k$, the
generalized scrambling time appears to follow the form
\begin{eqnarray}
t^* = v_B L + \Delta \sqrt{L} \, k.
\end{eqnarray}

The linear growth with $k$ is consistent with a linear growth of design order
(and therefore complexity) with circuit depth.  That is, we show that for the
values of $k$ studied here, there exists a regime beyond scrambling in local
random circuits where the wave function is definitively \emph{not} a (k/2)-design,
and the size of this regime appears to grow linearly with k.

\section{Discussion}

In this paper, we have studied in detail the quantum state complexity of wave
functions which are the output of local automaton circuits. These gates act very
simply on computational basis states, which allows us to simulate these circuit
efficiently using a classical computer. Despite this, when acting on an initial
product state with spins oriented perpendicular to the computational basis,
local random automaton circuits generate very complex highly entangled quantum
wave functions.  

We can quantify the complexity of the wave functions using tools from
quantum information theory.  We specifically relate the quantum state complexity
to the difficulty of distinguishing the wave function from the maximally mixed
state.   
We considered several different measurements in order to argue that
the ensemble of automaton wave functions form an approximate projective unitary
design of high order. Based on known connections between unitary designs and
quantum complexity, these results imply that automaton wave functions have a high
quantum state complexity. 

In the first section, we considered four basic measures of complexity.
First, we saw that in quantum automaton circuits, the distribution of bit-strings
(which result from many-qubit projective measurements in a basis orthogonal to
the computational basis) follows the well known `Porter-Thomas' distribution up to the
resolution of our numerics. Second, we characterized von Neumann entanglement
entropy across all possible bipartitions of the lattice and found that it is
always nearly exactly equal to the Page entropy. This is in contrast to Clifford
circuits, which show significant fluctuations between different partitions. We
also studied the level spacing distribution in the entanglement spectrum and saw
a convergence to the Gaussian unitary ensemble (GUE) Wigner-Dyson distribution,
a result that implies the reduced density matrix of automaton wave functions
behave like random matrices.  Finally, we studied the generalized k-th Renyi
entropies, which were shown in \cite{LiuLloyd} to be a direct probe of unitary design
complexity.  We found that for automaton wave functions, the k-th Renyi entropy
was exactly equal to the Haar random value for all values of $k$ we measured.
This suggests that design complexity in automaton wave functions will grow 
until the infinite order limit of the Renyi entropy is saturated, a condition
known as `max scrambling'. 

By every metric, the automaton circuits exhibit the same properties as a
wave function from high depth local random circuit composed of \emph{universal} basic gates. 
Such circuits are known to form an approximate polynomial unitary
design and therefore at sufficient depth approximate the fully Haar random wave
functions to arbitrary accuracy.  
All the above results suggest that fluctuations in automaton wave functions are
highly suppressed in the same way as in universal random circuits. In fact, the
bit-string distribution and generalized Renyi entropies follow the constraints
imposed for large k projective k-designs, implying that automaton wave functions
approximate at least the first $k$ moments of the fully random Haar measure. 
Further, convergence of the bit-string distribution to
the Porter-Thomas form and the entanglement level spacing to the Wigner-Dyson
distribution are often cited as key signatures of quantum chaos.  Throughout, we
compared the results for automaton circuits to those of Clifford circuits which
are known to form only a finite low order unitary design, even for high depth
circuits. This comparison serves to  highlight the difference between high and
low complexity states.  

In the second section, we studied the $2k$-point out-of-time-ordered
correlation (OTOC) functions.  These are a generalization of the popular 4-point
OTOC, which is known to characterize the onset of scrambling in quantum systems.
Unlike the previous set of measurements which require an
exponential effort to compute using our quantum Monte Carlo algorithm, the
generalized OTOC functions are capable of probing complexity beyond the
scrambling regime using computational resources which scale linearly with system
size. Therefore, we can use these as a tool to efficiently study complexity
growth in very large circuits. 

We first identified a set of recursively defined 2k-point OTOCs whose precise
form is physically motivated and which can be easily generalized to very high
values of $k$. We argue that in low complexity unitary circuits, such as Clifford circuits,
these higher order recursive OTOCs do not generically decay. In automaton
circuits, however, we find that at some time $t^*$ beyond the scrambling time
these correlation functions always decay to an exponentially small value. This
is both further evidence that automaton wave functions form projective unitary
designs and proof that generalized OTOCs can be practically used to probe
complexity beyond scrambling in large systems. 

Importantly, using automaton circuits, we were also able to study more generic
forms of out-of-time ordered correlation functions. By searching over thousands
of possible k-point OTOCs up to $k=16$, we were able to identify special
orderings of operators for which the average correlators remain nonzero for a far
longer time even than the recursive OTOCs discussed above. Measuring these
special OTOCs gives us an unprecedented ability to numerically characterize the
rate of complexity growth in large local random circuits. In particular, we
found that the scrambling time for the k-th order OTOC appears to increase linearly with
$k$. Notably, this result is consistent with the linear growth of complexity which is
conjectured in the literature. 
 
Taken together, all of the above results are very strong evidence
that the automaton circuits are capable of producing wave functions
with high quantum state complexity.  
Our results therefore imply that automaton circuits are a rare example of a
numerically tractable system which can generically simulate quantum chaotic dynamics
on a classical computer. We expect there to be
a wide variety of applications across a wide range of fields. Already, this
technique has been applied to a range of quantum circuit models in the context
of understanding quantum dynamics in systems with different symmetries
\cite{iadecola2020nonergodic,feldmeier2020anomalous,iaconis1, Chen_2020}. We predict that it will become a prominent
technique for studying generic
chaotic circuit dynamics and may be an important tool for understanding how the
growth of complexity beyond scrambling plays a role in thermalizing condensed
matter systems. There also exist obvious applications to quantum information
theory, where characterizing the complexity of local circuits is a key problem.
Beyond this, our work may be useful for practical quantum information processing
tasks such as randomized benchmarking \cite{PhysRevLett.106.180504,Wallman_2014}
and decoupling \cite{Szehr_2013}, where unitary designs play a key role.  The
ability of OTOCs to identify unique observables in high depth random
circuits which are fully scrambled may be useful experimentally for
characterizing the fidelity of non-Clifford quantum circuits. In high energy
physics, characterizing the rate of complexity growth in quantum systems is
conjectured to be related, through the holographic principle, to the
growth of the volume in the bulk geometry beyond the event horizon in black holes. Future work
in identifying the specific form of higher order OTOCs which best characterize
the complexity growth  may therefore shed important insight into these problems.

\begin{acknowledgments}
\noindent JI thanks Rahul Nandkishore, Xiao Chen, Itamar Kimchi, Roger Melko, and Zi-Wen Liu for useful discussions.
This material is based upon work supported by the Air Force Office of Scientific Research under award
number FA9550-20-1-0222. 
\end{acknowledgments}

\bibliography{base}

\end{document}